\documentclass[10pt]{article}
\usepackage{amsthm,amsmath}
\providecommand{\mcal}{\ensuremath{\mathcal}}

\newcommand{\transker}[1][\mcal{T}]{\ensuremath{\left.\left<q+\frac{v}{2}\right|#1\left|q-\frac{v}{2}\right>\right.}}
\newcommand{\kernel}[1]{\ensuremath{\left.\left<q\right|#1\left|q'\right>\right.}}
\newcommand{\kernell}[1]{\ensuremath{\left.\left<q'\right|#1\left|q\right>\right.}}
\newcommand{\tkernel}[1]{\ensuremath{\left.\left<t_a(q)\right|#1\left|t_a(q')\right>\right.}}

\newcommand{\dirac}[2]{\ensuremath{\left<#1\left|#2\right>\right.}}
\newcommand{\opr}[1]{\ensuremath{{\mathbf{\mathsf{#1}}}}}

\newcommand{\abs}[1]{\ensuremath{\left|#1\right|}}

\newcommand{\sign}[1]{\ensuremath{\mbox{sgn}\left(#1\right)}}

\newcommand{\hpossion}[2][H]{\ensuremath{\left\{#1,#2\right\}}}
\newcommand{\hliov}[1][H]{\ensuremath{\mathcal{L}_{#1}}}

\newcommand{\dom}[1]{\ensuremath{\mathcal{D}(\opr{#1})}}

\providecommand{\lterm}[2]{\ensuremath{\alpha_{#1,#2}^{(0)}}}
\newcommand{\order}[1]{\ensuremath{\mathcal{O}(\hbar^{#1})}}
\newcommand{\binomial}[2]{\ensuremath{{#1 \choose #2}}}
\newcommand{\expo}[1]{\ensuremath{\mbox{e}^{#1}}}

\begin{document}
\title{\textbf{Shouldn't there be an antithesis to quantization?}\thanks{E. A. Galapon {\it J. Math. Phys.} {\bf 45} 3180-3215 (2004)}}
\author{Eric A. Galapon \thanks{email: eric.galapon@up.edu.ph}\\Theoretical Physics Group, National Institute of Physics\\University of the Philippines, Diliman Quezon City\\1101 Philippines}
\maketitle
\begin{abstract}
We raise the possibility of developing a theory of constructing quantum dynamical observables independent from quantization and deriving classical dynamical observables from pure quantum mechanical consideration. We do so by giving a detailed quantum mechanical derivation of the classical time of arrival at arbitrary arrival points for a particle in one dimension. 
\end{abstract}

\section{Introduction}\label{intro}

Recently we raised the problem of deriving classical dynamical observables from pure quantum mechanical consideration, and thus the problem of constructing quantum observables with classical counterparts without quantization \cite{gal1}. Our motivations  have been to break the circularity of quantization when invoking the correspondence principle \cite{beltra,mac1,mac2}, and to sidestep the well-known existence of obstruction to quantization in important spaces like the Euclidean space \cite{got11,got12,got2,got21,got1,got0,gro,van}. The former motivation arises from the need for quantum mechanics to be internally coherent and autonomous from classical mechanics if quantum mechanics were the preponderant of the two mechanical theories. On the other hand, the later motivation arises from the need for observables to satisfy certain commutation relations in keeping with, say, the required evolution properties of the observables. Thus in \cite{gal1}  we have introduced the idea of {\it supraquantization}---the derivation of the quantum observable corresponding to a given classical observable without quantization, and the subsequent derivation of the classical observable from its quantum counterpart, as opposed to quantization which is the derivation of the quantum observable corresponding to a given classical observable by means of an associative mapping of the  scalar-valued observable to an operator-valued observable.

And to illustrate our point of supraquantization and to demonstrate the general insufficiency of prescriptive quantization---particularly the Weyl quantization \cite{wey}---to satisfy required commutator values, we outlined in \cite{gal1} without proof a formal quantum mechanical derivation of the local form of the classical time of arrival in the neighborhood of the origin. In this paper, we attempt to place our ealier results on a firm foundation. We do so by (1) developing the quantum mechanical framework suitable to the idea of supraquantization, and by (2) proving explicitly our earlier assertions made within the proposed framework. It is then the aim of this paper to give a quantum mechanical derivation of the classical observable
\begin{equation}\label{prob}
T_x(q,p)=-\mbox{sgn}(p) \sqrt{\frac{\mu}{2}}\int_x^q \frac{dq'}{\sqrt{H(q,p)-V(q')}},
\end{equation}
where $T_x(q,p)$ is the time of arrival of a particle at some point $x$, whose Hamiltonian is $H(q,p)$. We will do so within the rigged Hilbert space formulation of quantum mechanics \cite{raf1,raf2,raf3,bo1,bo2,gel1,gel2}

This paper is organized as follows. In Section-\ref{framework} we outline the quantum mechanical framework in rigged Hilbert space suitable for our purposes. In Section-\ref{quantsupra} we give a brief review of quantization and Weyl quantization in particular, and discuss the idea of supraquantization, and deal with the transition to the classical regime. In Section-\ref{classtoa} we summarize the classical solution to the classical time of arrival and introduce the concept of global and local time of arrivals. In Section-\ref{supra} we formulate our quantum mechanical approach to deriving the classical time of arrival at the origin from pure quantum mechanical consideration. In Section-\ref{examples}, we explicitly supraquantize the classical time of arrivals of the harmonic oscillator and the quartic oscillator. In Section-\ref{entireanalytic} we prove, within the limits stated therein, the general result for arbitrary entire analytic potentials that the time of arrival can be derived from the supraquantization developed in Section-\ref{supra}. In Section-\ref{arbitrary} we give the extension of our derivation for arbitrary arrival points. And in Section-\ref{conclusion}, we devote some discussion on the relationship between quantization and supraquantization.
 
In this paper, though we are concerned with the derivation of the corresponding quantum time of arrival operator, we will not delve into the important question whether one can extract quantum time of arrival distributions from the constructed operator \cite{mug222111,mug22211,mug2221,mug221,mug21,mug1,mug2,mug4,mug5,mug3}, nor its relevance in the quantum time problem \cite{gal2,gal3}. We leave these issues open in the mean time.

\section{The Quantum Mechanical Framework}\label{framework}

\subsection{Single Hilbert Space Quantum Mechanics}
In the generalized single Hilbert space formulation of quantum mechanics, to every quantum mechanical system is assigned a generally infinite dimensional Hilbert space $\mcal{H}$ over the complex field; and to every pure state corresponds to a ray in $\mcal{H}$; and to every observable corresponds to a generally maximally symmetric densely defined operator in $\mcal{H}$ \cite{bus,dav}.

If the system is closed or it does not react back to its environment, its evolution is governed by a one parameter unitary group, $U_t=\mbox{e}^{-\frac{i}{\hbar}\opr{H}t}$, where $\opr{H}$ is the system Hamiltonian. In Heisenberg representation where states are stationary, observables evolve according to
\begin{equation}\label{popo}
\opr{A}_t=e^{i\opr{H}t}\opr{A}e^{-i\opr{H}t},
\end{equation}
\begin{equation}\label{ooopo}
i\hbar\dot{\opr{A}}_t=\left[\opr{A}_t,\opr{H}\right],
\end{equation}
where (\ref{ooopo}) is the infinitesimal form of (\ref{popo}). If either $\opr{H}$ or $\opr{A}$ are unbounded, then equations (\ref{popo}) and (\ref{ooopo}) should be properly defined to give meaning to them. In particular, equation (\ref{popo}) holds for all times $t$ if the domain of $\opr{A}$ is invariant under  $e^{-i\opr{H}t}$ for all $t$. It is possible that the evolution equation and its infinitesimal form hold only in some countable subset of the time coordinate.

While the Hilbert space formulation is successful in describing much of the quantum mechanics we know, it is not sufficient in the sense that it does not accommodate the eigenfunctions of observables with pure continuous spectrum. It is in this context that extension of quantum mechanics in a rigged Hilbert space has been proposed. Moreover, it is within the rigging of $\mcal{H}$ that will allow us to further generalize observables to include operators that are not necessarily operators in the system Hilbert space. 

\subsection{Rigged Hilbert Space Extension}

Let $\mcal{H}$ be the system Hilbert space. A rigged Hilbert space (RHS) for $\mcal{H}$ is a triplet, called a Gel'fand triplet, $\Phi^{\times}\supset \mcal{H}\supset \Phi$, where $\Phi$ is a dense subspace of $\mcal{H}$, and  is a locally convex topological space and is complete with respect to its own topology;  on the other hand, $\Phi^{\times}$ is the space of all continuous linear functionals on $\Phi$: An element $F$ of $\Phi^{\times}$ assigns to every $\phi$ in $\Phi$ a complex number denoted by $\dirac{F}{\varphi}$ with the properties 
$\dirac{F}{a\phi_1+b\phi_2}=a\dirac{F}{\phi_1}+b\dirac{F}{\phi_2}$,
for every pair $\phi_1$ and $\phi_2$ in $\Phi$, and for every pair of complex numbers $a$ and $b$; and
$\lim_{n\to \infty}\dirac{F}{\phi_n}=0,
$ for every sequence $\phi_n$ converging to zero in $\phi$.

In extending quantum mechanics in a rigged Hilbert space, one has to specify a particular rigging. But how do we determine the necessary rigging? Our answer to this question is limited to what is relevant and useful to our present purposes. A natural choice is the one provided by the Hamiltonian of the system under consideration. Let $\opr{H}$ be the Hamiltonian and let its domain be $\dom{H}$. We choose $\Phi$ in such a way that $\Phi$ is a dense subset of $\dom{H}$, and that $\Phi$ is invariant under the Hamiltonian, i.e. $\opr{H}:\Phi\subseteq \Phi$. Our motivation in choosing this particular rigging is for us to be able to extend the quantum evolution of observables in the rigged Hilbert space, as we will define below. Once $\Phi$ is specified, its functional space $\Phi^{\times}$ is automatically determined.

Given the particular rigging of $\mcal{H}$ relative to $\Phi$, let $\mcal{O}$ be the set of all observables whose domains contain $\Phi$.  Let $\opr{A}$ be in $\mcal{O}$. We define two associated operators to $\opr{A}$: its rigged Hilbert space  {\it extension} $\opr{A}^{\times}$, and its rigged Hilbert space {\it reduction} $\opr{A}_{\times}$.
\begin{description}
\item[Rigged Hilbert Space Extension:] $\opr{A}^{\times}$ is the extension of $\opr{A}$ in the entire $\Phi^{\times}$, i.e. the operator $\opr{A}^{\times}:\Phi^{\times}\mapsto \Phi^{\times}$, such that
$\dirac{\opr{A}^{\times}\phi}{\varphi}=\dirac{\phi}{\opr{A}^{\dagger}\varphi}$
for all $\phi$ in $\Phi^{\times}$ and $\varphi$ in $\Phi$, where $\opr{A}^{\dagger}$ is the adjoint of $\opr{A}$ in $\mcal{H}$. 

\item[Rigged Hilbert Space Reduction:] $\opr{A}_{\times}$ is the reduction of $\opr{A}$ in $\Phi$, i.e. the operator $\opr{A}_{\times}\varphi=\opr{A}\varphi$ for all $\varphi$ in $\Phi$, such that there exists a uniquely associated functional $F_{\opr{A}}$ in $\Phi^{\times}$, for which $\opr{A}_{\times}\varphi=\dirac{F_{\opr{A}}}{\varphi}=\opr{A}\varphi$  for all $\varphi$ in $\Phi$. (See Appendix-1 for an example and to establish our notation.)
\end{description}
The rigged Hilbert space extension of $\opr{A}$ exists if $\opr{A}^{\dagger}$ is in $\mcal{O}$ and $\Phi$ is invariant under $\opr{A}^{\dagger}$, i.e. $\opr{A}^{\dagger}:\Phi\subseteq \Phi$. We emphasize that the definition of the rigged Hilbert space reduction of $\opr{A}$ requires the existence of the functional $F_{\opr{A}}$ in $\Phi^{\times}$ satisfying the stated condition. Of course $\opr{A}$ will always have a reduction in $\Phi$ by restricting its domain to $\Phi$, but it is not necessary that there is always an associated functional $F_{\opr{A}}$ in $\Phi^{\times}$. All throughout we will call $F_{\opr{A}}$ as the functional kernel of $\opr{A}_{\times}$.

Now eigenfunctions corresponding to the continuous part of the spectrum of an observable do not belong to the Hilbert space: they are not square integrable and the usual probabilistic interpretation of quantum mechanics fails to hold on them. But these acquire rigorous meaning within the context of a rigged Hilbert space as generalized eigenfunctions residing in the functional space $\Phi^{\times}$. If one can give physical significance to elements of $\Phi^{\times}$, it may also be possible to give physical significance to operators taking $\Phi$ into $\Phi^{\times}$, e.g. Hamiltonians with singular potentials in the configuration space, $V(q)\propto \delta(q-q_0)$. This motivates us to introduce the concept of generalized observable.
\begin{description}
\item[Generalized Observable:] Let ${F}$ be in $\Phi^{\times}$. If for all $\varphi$ in $\Phi$, ${F}(\varphi)=\dirac{{F}}{\varphi}$ is in $\Phi^{\times}$ and $\dirac{{F}(\varphi)}{\varphi}$ is real valued, we call ${F}$ to be the functional kernel of a generalized observable $\mcal{A}$.  The mapping $\dirac{F}{\cdot}:\Phi\mapsto \Phi^{\times}$ defines the generalized observable $\mcal{A}:\Phi\mapsto \Phi^{\times}$.
\end{description}
The real valuedness of $\dirac{{F}(\varphi)}{\varphi}$ is the generalized analogue of the symmetry condition in ordinary Hilbert space quantum mechanics, the numerical value of which is the expectation value of the generalized observable. Since we have the inclusion relation $\Phi^{\times}\supset \mcal{H}\supset \Phi$, the rigged Hilbert space reduction of ordinary quantum mechanical observables is a special class of generalized observables.  We say that $\mcal{A}$ has a Hilbert space projection if there exists a dense subspace $\mcal{D}$ of $\Phi$ such that $\dirac{{F}}{\varphi}$ is in $\mcal{H}$. Its Hilbert space projection is the closure of the operator $\opr{F}$ defined by $\dirac{{F}}{\varphi}=\opr{F}\varphi$ for all $\varphi$ in $\mcal{D}$. We emphasize that the properties of generalized observables are dictated by $\Phi$. For this reason we denote $\mcal{O}_{\Phi}$ to be the set of all generalized observables defined for a given $\Phi$.

Now we give the appropriate generalization of the evolution law for quantum observables. Let $\opr{H}$ be the system Hamiltonian, whose domain $\dom{H}$ contains $\Phi$. Let $\Phi^{\times}\supset \mcal{H} \supset \Phi$ be a particular rigging of $\mcal{H}$, where $\Phi$ is invariant under $\opr{H}$. Given $\opr{H}$, let $\opr{H}^{\times}$ be its RHS-extension. Let $\mcal{A}$ be a generalized observable in $\mcal{O}_{\Phi}$. If $\Phi$ is invariant under $U_t=e^{-i\opr{H}t}$ for all $t$, we then take $\mcal{A}$ to evolve according to
\begin{equation}\label{evoevo}
\mcal{A}_t=U_{-t}^{\times}\mcal{A} U_t^{\times},
\end{equation}
where $U_t^{\times}$ is the RHS-extension of $U_t$ in the entire $\Phi^{\times}$. Under the same assumption, the infinitesimal form of (\ref{evoevo}) is given by
\begin{equation}\label{evopa}
\frac{d\mcal{A}_t}{dt}=\frac{1}{i\hbar}\left[\mcal{A}_t, \opr{H}^{\times}\right].
\end{equation}
Equation (\ref{evoevo}) requires that $\mcal{A}$ evolves into a generalized observable. These equations reduce to the standard quantum evolution law when restricted to the Hilbert space. While equations (\ref{evoevo}) and (\ref{evopa}) hold for all $t$ under the assumption that $\Phi$ is invariant under $U_t$, they may still hold for some times, possibly countably infinite, even when $\Phi$ is not invariant under $U_t$ for  all $t$. 

\subsection{Quantum Mechanics in Configuration Space}\label{configspace}
For a spin less particle in the real line, the corresponding system Hilbert space is the space of Lebesgue square integrable functions over the real line, $\mcal{H}=L^2(\Re,dq)$. We assume that the particle is under the influence of an everywhere infinitely differentiable (real valued) potential, i.e of type $C^{\infty}(\Re)$. The formal Hamiltonian
\begin{equation}
\opr{H}=-\frac{\hbar^2}{2\mu}\frac{d^2}{dq^2}+V(q)
\end{equation}
can be assigned a dense domain $\dom{H}$ in which it is essentially self-adjoint. The given Hamiltonian allows several possible riggings of the Hilbert space. We choose the rigging $\Phi^{\times}\supset \mcal{H} \supset \Phi$, where $\Phi$ is the space of infinitely differentiable complex valued functions with compact support in the real line, and $\Phi^{\times}$ its corresponding functional space. Since $V(q)$ is $C^{\infty}(\Re)$, $\Phi$ is invariant under $\opr{H}$. We note that $\Phi$ is tight enough to allow a larger $\Phi^{\times}$.

The convergence to zero of a sequence in $\Phi$ is defined as follows. A sequence $\varphi_n$ in $\Phi$ converges to zero in $\Phi$ if all these functions vanish outside a certain fixed bounded region, the same for all of them, and converge uniformly to zero in the usual sense together with their derivatives of any order \cite{gel1,gel2}.

With our chosen rigging, the rigged Hilbert space reduction $\opr{A}_{\times}$ of an operator $\opr{A}$ in $\mcal{H}$ with domain containing $\Phi$ assumes the familiar form
\begin{equation}\label{daniel}
\left(\opr{A}_{\times}\varphi\right)\!(q)=\dirac{F_{\opr{A}}(q)}{\varphi}=\int_{\Re}\kernel{\opr{A}_{\times}}\varphi(q')\, dq', 
\end{equation}
where the functional kernel $F_{\opr{A}}$ is the complex conjugate of the well-known configuration matrix elements of the operators. Generalized observables, which includes the RHS-reductions of operators in $\mcal{H}$, assume the similar form
\begin{equation}\label{mary}
\left(\mcal{A}\varphi\right)\!(q)=\int_{\Re}\kernel{\mcal{A}}\varphi(q')\, dq', 
\end{equation}
where the integrations in equations (\ref{daniel}) and (\ref{mary}) are understood to be in distributional sense, in particular symbolic when the integrand is singular, e.g. the Dirac delta function. We note that the functional kernel $\kernel{\mcal{A}}$ must be symmetric, i.e. $\kernel{\mcal{A}}=\kernell{\mcal{A}}^*$, in order to ensure the real valuedness of the expectation value of $\mcal{A}$ in $\Phi$. 

\section{Quantization, Supraquantization, and the Transition to the Classical Regime}\label{quantsupra}
\subsection{Quantization and Weyl-Quantization}
Let $f$ be a classical observable, a real valued function $f(q,p)$ in the phase space. 
The problem of quantization is to derive the quantum counterpart of $f$ by some associative mapping $Q$ of the real-valued function $f(q,p)$ to a maximally symmetric operator $\opr{F}$ in the system Hilbert space $\mcal{H}$, i.e. $Q(f)\mapsto \opr{F}$. A paramount requirement, aside from other requirements, of quantization is that the possion bracket of two (classical) observables quantizes into the commutator of the separately quantized observables, in particular $Q(\left\{f,g\right\})= (i\hbar)^{-1}\left[Q(f),Q(g)\right]$ (for a complete discussion on the requirements of quantization, see References-\cite{got2,got0}).

One of the earliest prescription, which has become the starting point of other quantization schemes, is the Weyl quantization $Q_W$. In the language of the framework outlined above, Weyl quantization is the bijective mapping of $f$ into some functional of a particular rigging $\Phi^{\times}\supset \mcal{H} \supset \Phi$ of the system Hilbert space, i.e. $Q_W:f\mapsto F\in \Phi^{\times}$ such that $F(\cdot)$ is a generalized observable with a non-trivial Hilbert space projection. The rigging of $\mcal{H}$ which we have required above is appropriate for Weyl's quantization. Now $Q_W$ is defined by the mapping
\begin{equation}\label{kerker}
Q_W:f\mapsto \Phi^{\times}\ni F^*=\kernel{\opr{F}_{\times}}=\frac{1}{2\pi \hbar}\int_{-\infty}^{\infty} f\!\left(\frac{q+q'}{2},p\right) \exp\left[\frac{i}{\hbar}(q-q')p\right]\, dp,
\end{equation}
where the integration is done in the distributional sense. In the standard formulation, it is assumed, though it is {\it not} guaranteed, that $F(\cdot)$ has a Hilbert space projection, so that $F(\cdot)$ is the rigged Hilbert reduction $\opr{F}_{\times}$ of a uniquely associated operator $\opr{F}$ in $\mcal{H}$. The operator $\opr{F}$ is  the closure of $\opr{F}_{\times}$ in $\mcal{H}$. 

Now suppose that $\opr{F}$ is the Hilbert operator corresponding to the classical operator $f$ derived by quantization. The classical observable is recovered by mere inversion of the process. In this case one has to determine the rigged Hilbert space reduction $\opr{F}_{\times}$ of the operator $\opr{F}$ and consequently determine the functional kernel corresponding to $\opr{F}_{\times}$. Given the functional kernel, the classical observable is recovered by means of the inverse Fourier transform
\begin{equation}\label{gentransitionweyl}
f(q,p)= \int\limits_{-\infty}^{\infty}\transker[\opr{F}_{\times}]\,\exp\left(-i\frac{v\,p}{\hbar}\right) \,dv.
\end{equation}
In this expression, taking the limit as $\hbar$ approaches zero is not required, it being  just the inverse of the prescribed Weyl-quantization. 

Quantization, however, is circular when invoking the correspondence principle; and this is already evident for the Weyl quantization. Moreover, there is a well-known obstruction to quantization in Euclidean space (and other spaces) which says that no quantization exists that satisfies the poission-bracket-commutator correspondence requirement for all observables \cite{got11,got12,got2,got21,got1,got0,gro,van}. This is unsatisfactory because the said correspondence is necessary, for example, in ensuring that required evolution properties of a certain class of observables are satisfied. This handicap of quantization will be explicitly demonstrated for the Weyl quantization. 

If we wish to break the inherent circularity of quantization and to sidestep the obstruction to quantization, we must then find an alternative platform upon which we can construct quantum observables without quantization, which can further allow us to derive the corresponding classical observable. It is here that the idea of supraquantization comes in.

\subsection{Supraquantization}
The idea behind supraquantization---the construction of quantum observables without quantization and the subsequent quantum mechanical derivation of its classical counterpart---is not entirely new. 

It has its origin in Mackey's earlier effort of restoring the autonomy of quantum mechanics from classical mechanics \cite{mac1,mac2}. We recall that the quantization of free particle in one dimension is accomplished by promoting its position and momentum into operators and their Poission bracket into commutator, and the energy into the Hamiltonian operator.  Mackey's work obviates these quantization prescriptions by starting not from the classical description but from the axioms of quantum mechanics and the property of free space. Starting from the basic axiom that the proposition for the location of the particle in different volume elements are compatible and the fundamental homogeneity of free space, one derives the position and the momentum operators together with the canonical commutation relation they satisfy. On the other hand, requiring Galilean invariance in the lattice of propositions, one derives the free quantum Hamiltonian (see also Reference-\cite{jau}). Mackey's work provides an excellent example of the existence of more than one solution to the derivation of the quantum image of a given set of classical observables. By definition Mackey's construction of the position and momentum operators, together with their algebra, is a supraquantization of the classical position and momentum, and their possion algebra. 

It maybe that quantization and supraquantization yield the same results, like the position and momentum for the free particle, but they are unmistakably distinct. Quantization presupposes classical mechanics, while supraquantization upholds the autonomy of quantum mechanics; the former introduces circularity when invoking the correspondence principle, while the latter sanctions the correspondence principle as a legitimate consequence of the acknowledged preponderance of quantum mechanics over classical mechanics. In both methods of obtaining quantum observables,  the classical observable plays two different roles. In quantization, it is the starting point; in supraquantization, it is a boundary condition. The correspondence principle requires that if a quantum observable corresponds to a classical observable, then the former should reduce to the latter in the limit of vanishing $\hbar$. Then if supraquantization gives the correct quantum observable, then that observable should approach its classical limit. As a consequence of the role of the classical observable as a boundary condition, supraquantization breaks the vicious
circle inherent in the quantization procedure.

But how do we construct quantum observables corresponding to a given classical observable without quantization? The observable may be constructed by appealing to the {\it postulated properties of the observable under consideration}, or to the {\it postulated physical properties of the universe}, or to the {\it axioms of quantum mechanics}, or to {\it any combination of these}. Mackey's construction of the quantum position and momentum observables without quantization proceeds from the homogeneity of free space (assumed property of free space) and the commutativity of propositions for the location of a free particle (axiom of quantum mechanics). 

For a specific class of classical observables, the required supraquantization may be accomplished, in addition to the aforementioned method, by referring to one of the members of the class and employing a transfer principle to the rest. The transfer principle can be expressed as follows:
\begin{description}
\item[Transfer Principle]: Each element of a class of observables shares a common set of properties with the rest of its class such that when a particular property is identified for a specific element of the class that property can be transferred to the rest of the class without discrimination.
\end{description}
It is the central problem of supraquantization to determine this set of properties shared by the class of observables under consideration, together with the appropriate axioms of mechanics to impose. Obviously supraquantization treats each class of observables on a case to case basis, in contrast to quantization which gives a single rule of association between classical and quantum observables.

But how do we approach the classical limit? We can treat quantization as a first order approximation, especially in those cases where obstruction to quantization occurs, e.g. in Euclidean space, and treat its classical limit as the starting point. This is reasonable because quantization has been successful in cases where consistency is preserved. So for observables defined in $\mcal{H}=L^2[\Re,dq]$ or for generalized observables in a particular rigging of $\mcal{H}$, the transition to the classical regime is still given by equation (\ref{gentransitionweyl}) only that one now has to specifically project the observable into the $\hbar=0$ or $\hbar=\delta$ regime. This is so as some orders of $\hbar$ now generally appear in equation (\ref{gentransitionweyl}). The appearance of terms in some orders of $\hbar$ indicates the failure of quantization to consistently satisfy the required commutator values, at least in Weyl's quantization. Thus for all generalized observables $\mcal{A}$ definable relative to $\Phi$ with functional kernel $\kernel{{\mcal{A}_{\times}}}^*$, the classical limit of $\mcal{A}$ is given by 
\begin{equation}\label{gentransition}
f(q,p)=\lim_{\hbar\to 0} \int\limits_{-\infty}^{\infty}\transker[\mcal{F}]\,\exp\left(-i\frac{v\,p}{\hbar}\right) \,dv
\end{equation}
whenever the limit exists (equation (\ref{gentransition}) has already been in used, see \cite{deg,omn}. The vanishing of $\hbar$ in the above expression is the statement that classical mechanics is the projection of quantum mechanics. Because classical mechanics is a projection, there is no bijection from classical to quantum mechanics, except in those cases where the results of quantization and supraquantization agree. This will be made clear when we consider the supraquantization of the classical time of arrival.

\section{The Observable on Case: The Classical Time of Arrival}\label{classtoa}
Consider a particle with mass $\mu$ in one dimension whose Hamiltonian is $H(q,p)$. If at $t=0$ the particle is at the point $(q,p)$ in the phase space, the time $t=T_x$ that the particle will arrive at the point $q(t=T_x)=x$ is given by 
\begin{equation}\label{classpass}
		T_x(q,p)=-\sign{p} \sqrt{\frac{\mu}{2}}\int_x^q \frac{dq'}{\sqrt{H(q,p)-V(q')}},
\end{equation}
derived by inverting the (classical) equations of motion. For a given energy the region $\Omega=\Omega_q\times \Omega_p$ in the phase space in which equation (\ref{classpass}) is finite and real valued is the classically accessible region to the particle for a given arrival point $x$. An important property of $T_x(q,p)$ is that it evolves according to
\begin{equation}
\frac{dT_x(q,p)}{dt}=-1.
\end{equation}
This property will be important to us in the supraquantization of the time of arrival.

It is the goal of this paper to show that the time of arrival (\ref{classpass}) for entire analytic potentials can be derived within the quantum mechanical framework we have just outlined above. This we will accomplish by constructing the generalized quantum observable corresponding to $T_x(q,p)$ by supraquantization to be developed below. Before we can proceed, we must recognize that $T_x(q,p)$ is only an observable in the region of the phase space accessible to the particle. Supraquantization of $T_x(q,p)$ then must be restricted to these accessible regions. We then proceed by developing a local form of $T_x(q,p)$, i.e. an equivalent expression for $T_x(q,p)$ in some neighborhood of $\Omega_q$, which can be assured to be finite and real valued, thus an observable. It is this local form, which we shall refer to as the local time of arrival, that we will supraquantize and show to be derivable quantum mechanically. The time of arrivals for the rest of the accessible regions are then derived by simple analytic continuation of the local time of arrival. In the following section we develop the local expression for the time of arrival for arbitrary arrival $x$.

\subsection{The Local Time of Arrival}
Given the Hamiltonian $H=\frac{1}{2\mu}p^2+V(q,p)$, let us consider all real valued functions, $T(q,p)$, in the phase space which is canonically conjugate with the Hamiltonian, i.e.
\begin{equation}
\left\{H(q,p),T(q,p)\right\}=1
\end{equation}
where $\{,\}$ is the Possion bracket. The time of arrival at some specified point is one such phase space function. Out of all those $T(q,p)$'s conjugate with $H(q,p)$, let us consider those that can be parametrized by $x'$ and $h$, where $x'$ is in the configuration axis and $h$ is a fixed function of $p$ alone. We denote these by $T_h^{x'}(q,p)$. The parameters $x'$ and $h$ are defined as follows. 
Let $K=\frac{1}{2\mu}p^2$ be the kinetic energy, and $\hliov[K]$ be the kinetic energy Liovillian operator defined by $\hliov[K]\cdot g=\{K,g\}=-\mu^{-1}\,p \, \partial_q g$. The pair of parameters $x'$ and $h$ fixes the inverse of $\hliov[K]$, $\hliov[K]^{-1}$, as follows
\begin{equation}\label{5inverse}
\hliov[K]^{-1}\cdot f(q,p)=-\frac{\mu}{p} \int_{x'}^{q} f(q',p)\, dq' +h(p).
\end{equation}
In other words, $x'$ and $h$ define the domain of $\hliov[K]$ such that the inverse $\hliov[K]^{-1}$ can be unambiguously defined.

Now given $x'$ and $h$ we construct $T_h^{x'}$ by the following prescription. Since $\hpossion{T_h^{x'}}=\hliov\cdot T_g (q,p)=1$, we express $T_h^{-1}$ in the following form
\begin{eqnarray}
		T_h^{x'} (q,p)=\hliov^{-1}\cdot 1			  =\frac{1}{\hliov[K]+\hliov[V]}\cdot 1\label{form1}
\end{eqnarray}
where $K$ and $V$ are the kinetic and potential energy parts of the Hamiltonian, respectively. Geometric expansion of equation (\ref{form1}) yields
\begin{equation}\label{expansion}
		T_g^{x'} (q,p)=\hliov[K]^{-1}\cdot 1 
				- \hliov[K]^{-1}\cdot \hliov[V]\cdot \hliov[K]^{-1}\cdot 1 
				+ \hliov[K]^{-1}\cdot \hliov[V]\cdot \hliov[K]^{-1}\cdot \hliov[V]\cdot 
						\hliov[K]^{-1}\cdot 1\cdots
\end{equation}
where $\hliov[K]^{-1}$ is defined by equation (\ref{5inverse}). Assuming that there is a neighborhood in the phase space such that the right hand side of (\ref{expansion}) converges, equation (\ref{expansion}) can be written in series form
\begin{equation}\label{series}
		T_g^{x'} (q,p)=\sum_{k=0}^{\infty}(-1)^k\,T_k (q,p,x')
\end{equation}
where the $T_k (q,p,x')$'s satisfy the recurrence relation
\begin{equation}\label{recurrence}
		T_0 (q,p,x') = \hliov[K]^{-1}\cdot 1,\;\;\;
		T_k (q,p,x')=\hliov[K]^{-1}\cdot\hliov[V]\cdot T_{k-1}(q,p)
\end{equation}
The system of recurrence relation (\ref{recurrence}) is equivalent to the recurrence relation $\hpossion[K]{T_k}=\hpossion[V]{T_{k-1}}$, subject to the boundary condition $\hpossion[K]{T_0}=1$. For a given $x'$ and $h$, equations (\ref{recurrence}) assume the explicit forms,
\begin{equation}\label{ker}
T_0 (q,p,x')=-\frac{\mu}{p}(q-x')+h(p)
\end{equation}
\begin{equation}\label{cow}
		T_k (q,p,x')=-\frac{\mu}{p}\int_{x'}^q \left(\frac{\partial V}{\partial q'}
			\frac{\partial T_{k-1}}{\partial p}-\frac{\partial V}{\partial p}
			\frac{\partial T_{k-1}}{\partial q'}\right)\,dq' + h(p)
\end{equation}
For autonomous Hamiltonian systems, i.e. $V=V(q)$, equation (\ref{cow}) reduces to
\begin{equation}\label{cowcow}
		T_k (q,p,x')=-\frac{\mu}{p}\int_{x'}^q \frac{\partial V}{\partial q'}
				\frac{\partial T_{k-1}}{\partial p}\,dq' + h(p).
\end{equation}
Of course equation (\ref{series}) need not converge. However for some conditions to be stated below it converges to the time of arrival in some neighborhood.

Now let $h=0$, $p\neq 0$ and $V(q)$ be continuous at $q$ where $q$ is an interior point of $\Omega_q$. Then there exists a neighborhood of $q$, $\omega_q\subseteq\Omega_q$, determined by the neighborhood $\abs{V(q)-V(q')}<K_{\epsilon}\leq \frac{p^2}{2\mu}$ such that for every $x\in \omega_q$, $T_0^x$ converges absolutely and uniformly to the classical time of arrival $t_x$.

We prove this assertion as follows. With $h=0$, equations (\ref{ker}) and (\ref{cowcow}) reduce to
\begin{equation}\label{inini}
			T_0(q,p;x)=-\mu \frac{(q-x)}{p},
\end{equation}
\begin{equation}\label{mowcow}
		T_k (q,p,x')=-\frac{\mu}{p}\int_{x}^q \frac{\partial V}{\partial q'}
				\frac{\partial T_{k-1}}{\partial p}\,dq'.
\end{equation}
Using equation (\ref{mowcow}) with $x'=x$, and using (\ref{inini}) as the initial value, the first few terms in equation (\ref{expansion}) can be evaluated to aid us to infer that the $k$-th iterate, $T_k$, in equation (\ref{series}) is given by
\begin{equation}\label{okay}
	T_k (q,p;x)=-\frac{(2k-1)!!}{k!} \frac{\mu^{k+1}}{p^{2k+1}}\int_x^q \left(V(q)-V(q')\right)^k\, dq'.
\end{equation}
We prove equation (\ref{okay}) by induction. Shifting index $k\rightarrow (k-1)$ in (\ref{okay}) to get $T_{k-1}$ and substituting $T_{k-1}$ back into equation (\ref{cowcow}), we have
\begin{eqnarray}
T_k(q,p;x)&=&-\frac{\mu}{p} \int_x^q \frac{\partial V}{\partial q'}
					\frac{\partial T_{k-1}}{\partial p} \, dq' \nonumber\\
&=&-\frac{(2k-3)!!}{(k-1)!}(2k-1) \frac{\mu^{k+1}}{p^{2k+1}} \int_x^q \frac{\partial V}{\partial q'} \int_x^{q'}\left(V(q')-V(q'')\right)^{k-1}\, dq''\label{5mawmaw}
\end{eqnarray}
Successive integration by parts evaluates the double integration into
\begin{eqnarray}
\lefteqn{\int_x^q \frac{\partial V}{\partial q'} \int_x^{q'} (V(q')-V(q''))^{k -1} \, dq''} 
\hspace{25.0 mm}\nonumber\\
&=& V(q)\int_x^q \left(V(q)-V(q')\right)^{k-1}\, dq'\nonumber \\
& & -\frac{(k-1)}{2}\int_x^q \frac{d V^2}{dq'} \int_x^{q'} \left(V(q')-V(q'')\right)^{k-2}\, dq'' dq' \nonumber \\
&=& \int_x^q \sum_{j=0}^{k-1} (-1)^j \frac{(k-1)!}{(k-1-j)!(j+1)!} V^{j+1}(q) \left(V(q)-V(q')\right)^{k-1-j} \nonumber \\
& & \hspace{20mm}+ \frac{(-1)^k}{k} \int_x^q V^k(q')\, dq' \nonumber\\
&=&\frac{1}{k}\int_x^q \left(V(q)-V(q')\right)^k\, dq' \label{finfin}.
\end{eqnarray}
Substituting equation (\ref{finfin}) back into eqn (\ref{5mawmaw}), we get 
\begin{equation}\label{pino}
T_k(q,p;x) = -\frac{(2k-3)!!}{(k-1)!}\frac{(2k-1)}{k} \int_x^q \left(V(q)-V(q')\right)^k\, dq',
\end{equation}
which reproduces and validates equation (\ref{okay}). Equation (\ref{series}) then reduces to the form 
\begin{equation}\label{5deded}
T_0^x=-\sum_{k=0}^{\infty}(-1)^k \frac{(2k-1)!!}{k!} \frac{\mu^{k}}{p^{2k+1}}\int_x^q (V(q)-V(q'))^k \,dq'.
\end{equation}
Of course equation (\ref{5deded}) does not necessarily converge. We next tackle this convergence issue.

Let us consider the neighborhood of $V(q)$ given by $\abs{V(q)-V(q')}<K_{\epsilon}$ for some $K_{\epsilon}\leq p^2(2\mu)^{-1}$. By the continuity of $V$ at $q$, there exists a neighborhood of $q$, $\omega_q$, completely determined by the neighborhood $\abs{V(q)-V(q')}<K_{\epsilon}$. Now let $x $ be in $\omega_q$ and consider the closed interval $\Delta=[q,x]$ which is contained in $\omega_q$. Because $V$ is continuous in the neighborhood $\omega_q$, it is likewise continuous in $\Delta$. This implies that, for a fixed $q$, $\abs{V(q)-V(q')}$ as function of $q'$ in the interval $\Delta$ possesses an absolute maximum $M_q$. Thus we have the inequality,
\begin{equation}
 \abs{\sum_{k=0}^{\infty}(-1)^k \frac{(2k-1)!!}{k!} \frac{\mu^{k}}{p^{2k}}\int_q^x (V(q)-V(q'))^k \,dq'}\leq \sum_{k=0}^{\infty}\frac{(2k-1)!!}{k!} \frac{\mu^{k}}{p^{2k}}M_q^k\, (x-q) \label{fog}
\end{equation}
The right hand side of inequality (\ref{fog}) converges absolutely if and only if $\mu\,p^{-2}M_q<\frac{1}{2}$. Because $M_q<K_{\epsilon}\,p^2(2\mu)^{-1}$, the right hand side of inequality (\ref{fog}) absolutely converges. This implies that equation (\ref{series}) converges absolutely for every $x$ in $\omega_q$.  The absolute convergence of the right hand side of inequality (\ref{fog}) also implies the uniform converge of equation (\ref{series}) because we can always replace $(x-q)$ by $l$ in (\ref{fog}) where $l$ is the length of any interval containing $\Delta$.

To show that $T_g^x$ converges absolutely and uniformly to $T_x$, we must show that the indicated integration in each term of the series can be factored out. This happens when the series $\sum_{k=0}^{\infty}(-1)^k (2k-1)!!(k!)^{-1}\mu^{k}\,{p^{-2k}}(V(q)-V(q'))^k 
$ converges uniformly for a fixed $q$ and for every $q'$ in $\Delta$. This, in fact, is ensured by the absolute convergence of (\ref{fog}). Pulling the integral out in equation  (\ref{5deded}), we arrive at
\begin{eqnarray}
T_0^x (q,p)&=&-\int_x^q \left(\sum_{k=0}^{\infty}(-1)^k \frac{(2k-1)!!}{k!}\frac{\mu^{k+1}}{p^{2k+1}}(V(q)-V(q'))^k \right)\,dq' \nonumber \\
&=&-\frac{\mu}{p}\int_x^q \left(\sqrt{1+\frac{2\mu(V(q)-V(q'))}{p^2}}\right)^{-\frac{1}{2}}\label{opp}\, dq'
\end{eqnarray}
Writing $p=\sign{p}\sqrt{\abs{p}^2}$ in equation (\ref{opp}) finally yields
\begin{equation}
T_g^x (q,p)=-\sign{p}\sqrt{\frac{\mu}{2}}\int_x^q \frac{dq'}{\sqrt{H(q,p)-V(q')}},
\end{equation}
which is just the time of arrival at $x$. Thus $T_g^x (q,p)=t_x(q,p)$ in $\omega\subset\Omega$. 

Because $T_x(q,p)$ holds in the entire $\Omega$ by definition and $T_0^x (q,p)$ holds only in some local neighborhood  $\omega_q$ of $\Omega_q$, we have the inclusion $T_0^x (q,p)\subset T_x (q,p)$; that is, $T_x (q,p)$ is the analytic continuation of $T_0^x (q,p)$ in $\Omega\setminus\omega$. For this reason we refer to $T_0^x (q,p)$ as the {\it local time of arrival} at $x$, and $T_x(q,p)$ as the global time of arrival. As we have mentioned above it is the local form or the local time of arrival that we will derive quantum mechanically, so that the global time of arrival is only derived by extension.

\section{Supraquantization of the Classical Time of\\Arrival}\label{supra}

\subsection{The Problem}

Let $\mcal{H}$ be the system Hilbert space and $\opr{H}=\frac{1}{2\mu}\opr{p}^2+V(\opr{q})$ be its Hamiltonian where $V(q)$ is $C^{\infty}(\Re)$. Following Section-\ref{configspace}, the rigging of $\mcal{H}$ is  $\Phi^{\times}\supset \mcal{H} \supset \Phi$, where $\Phi$ is the fundamental space of infinitely differentiable complex valued functions with compact support in $\Re$, and where $\Phi^{\times}$ is the corresponding functional space for $\Phi$.

The rigged Hilbert space extension of the Hamiltonian $\opr{H}$ is then explicitly given by
\begin{equation} \label{Hamiltonian}
    \opr{H}^{\times}\phi=-\frac{\hbar^2}{2\mu} \frac{d^2\phi}{dq^2}+V(q)\phi \;\;\;\mbox{for all} \;\;\phi \in \Phi^{\times}.
\end{equation}
In this paper we assume that the potential $V(q)$ is at most entire analytic in $q$, i.e. represented by an everywhere convergent power series in $q$. The entire analicity of $V(q)$ is consistent with our requirement that $\Phi$ is invariant under the action of the Hamiltonian.

Given the Hamiltonian $\opr{H}$, our problem is to construct the corresponding generalized time of arrival operator $\mcal{T}$ consistent with the correspondence principle: $\mcal{T}$ reducing to the local time of arrival in the classical limit. The operator $\mcal{T}$ is by hypothesis a generalized observable relative to the rigging provided by $\Phi$, i.e. the operator $\mcal{T}:\Phi\mapsto \Phi^{\times}$. This operator is then uniquely associated with a functional kernel $F_{\mcal{T}}$ defined by $\mcal{T}:\Phi=\dirac{F_{\mcal{T}}}{\cdot}:\Phi$. Explicitly 
\begin{equation} \label{fretimo}
 \left(\mcal{T}\varphi\right)\!(q)=\int_{-\infty}^{\infty}\kernel{\mcal{T}}\varphi(q')\,dq',
\end{equation}
where $\kernel{\mcal{T}}^*$ is the functional kernel $F_{\mcal{T}}$. As a generalized observable,  the functional kernel must be symmetric, i.e. $\kernel{\mcal{T}}=\kernell{\mcal{T}}^{*}$. Moreover, since $\mcal{T}$ is the quantum counterpart of the local time of arrival, it has to be that the classical local time of arrival operator is recovered by means of equation (\ref{gentransition}), specifically
\begin{equation}\label{transition}
t_0(q,p)=\lim_{\hbar\to 0} \int\limits_{-\infty}^{\infty}\transker\,\exp\left(-i\frac{v\,p}{\hbar}\right) \,dv.
\end{equation}
The problem of constructing $\mcal{T}$ then reduces to the problem of determining its functional kernel $\kernel{\mcal{T}}$. It is now the problem of supraquantization to determine $\kernel{\mcal{T}}$ without appealing to quantization. (We leave the problem whether $\mcal{T}$ has Hilbert space projection or none open, a problem relevant to the question whether time of arrival distributions can be extracted from $\mcal{T}$.)

\subsection{The Construction of Solution}
But how do we determine the kernel $\kernel{\mcal{T}}$ without resorting  to quantization? We accomplish this in two steps. First is by identifying the property of $\mcal{T}$ and implementing this property through the appropriate axiom of quantum mechanics. Being a time of arrival operator, it must at least evolve according to 
\begin{equation}\label{okum}
\frac{d\mcal{T}}{dt}=-\mcal{I}
\end{equation}
in which $\mcal{I}$ is the identity in $\Phi$. We note that it is not necessary that the above evolution law holds for all $t$. Fortunately, it is sufficient for us to require the condition $\dot{\mcal{T}}(0)=-\mcal{I}$, or $\mcal{T}$ evolves according equation (\ref{okum}) in the neighborhood of $t=0$. This is always satisfied because $\Phi$ is assumed to be invariant under the action of $\mcal{H}$. Imposing this on equation (\ref{evopa}), we arrive at the canonical commutation relation
\begin{equation} \label{condition}
    \dirac{\phi}{[\opr{H^{\times}},\mcal{T}]\varphi}=i\hbar \dirac{\phi}{\varphi}
\end{equation}
satisfied by the Hamiltonian and the time  of arrival operator, for all $\phi, \varphi \in \Phi$. Equation (\ref{condition}) is the basic condition satisfied by
$\mcal{T}$ but it is not sufficient to completely determine
$\mcal{T}$.

The second step is by employing a kind of transfer principle we mentioned earlier.  We hypothesize that each element of a class of time of arrival observables shares a common set of properties with the rest of its class such that when a particular property is identified for a specific element of the class that property can be transferred to the rest of the class without discrimination.

We exploit this in determining the kernel $\kernel{\mcal{T}}$ by solving  the simplest in the class of time of arrival observables, the free particle. We start by recalling that the
free particle is Galilean invariant, a consequence of the homogeneity of free space. It will not matter then where we place the origin. This implies that the commutation relation
(\ref{condition}) holds independent of the choice of origin. Because of this and because the free Hamiltonian is Galilean invariant, we require that the time kernel for the free particle
must itself be Galilean invariant. Specifically if $t_a$ is translation by $a$, i.e. $t_a(q)=q+a$ and if $\kernel{\mcal{T}}$ is the free particle kernel, then the translated free time of arrival operator $\mcal{T}_a=\int dq\,\tkernel{\mcal{T}}$ must still satisfy equation (\ref{condition}). In addition to translational invariance, $\kernel{\mcal{T}}$ must be symmetric, and it must be chosen such that equation (\ref{condition}) is
satisfied given the free Hamiltonian $\opr{H}\phi=-\hbar^2(2\mu)^{-1}\phi''$, and it must reproduce the free time of arrival at the origin via equation (\ref{transition}).  A solution satisfying all these requirements is given by
\begin{equation}\label{freefree}
\kernel{\mcal{T}}=\frac{\mu}{i\,4\hbar} (q+q')\, \mbox{sgn}(q-q').
\end{equation}
We note though that (\ref{freefree}) is not unique. The kernel $\hbar^{-1}\mu \left|a-a'\right|$ is dimensionally consistent with (\ref{freefree}) and it is Galilean invariant and it commutes with the free Hamiltonian in the entire $\Phi$. Moreover it vanishes in the classical limit. Then real factors of it can be added to (\ref{freefree}) without sacrificing any of the required properties of the free particle kernel. However, $\hbar^{-1}\mu \left|a-a'\right|$ arises only because of Galilean invariance which is an exclusive property of the free particle.  Since we are aiming at exploiting the assumed transfer principle, we can not carry it over to the rest of its class. 

Having solved the free particle kernel, we proceed in implementing the  transfer principle. We hypothesize that all time kernels assume the same form. Thus, from equation (\ref{freefree}), we assume that the time kernel is given by 
\begin{equation} \label{timekernel}
    \kernel{\mcal{T}}=\frac{\mu}{i\,\hbar}\; T(q,q')\, \mbox{sgn}(q-q')
\end{equation}
where $T(q,q')$ depends on the given Hamiltonian. Inferring from the free particle time kernel, we require that $T(q,q')$ be real valued,  symmetric, $T(q,q')=T(q',q)$, and analytic. We determine $T(q,q')$ by imposing condition (\ref{condition}) on $\mcal{T}$.
Substituting equation (\ref{timekernel}) back into the left hand side of equation (\ref{condition}) and performing two successive integration by parts, we arrive at 
\begin{eqnarray}
\lefteqn{\dirac{\phi}{[\opr{H}^{\times},\mcal{T}]\varphi}=i\hbar\!\int\limits_{\Sigma}\!\phi^{*}(q)\!\left(\frac{d T(q,q)}{dq} +  \frac{\partial T(q',q')}{\partial q} + \frac{\partial T(q,q)}{\partial q'}\right)\! \varphi(q)\,dq}\nonumber \\
& &\hspace{-4.5mm} -i\frac{\mu}{\hbar}\int\limits_{\Sigma} \phi^{*}(q)\left[\left(-\frac{\hbar^2}{2\mu} 		\frac{\partial^2}{\partial q^2} + V(q)\right)T(q,q')- \left(-\frac{\hbar^2}{2\mu}\frac{\partial^2}{\partial {q'}^2} 		+V(q')\right)T(q,q')\right]\nonumber \\
		& &\hspace{5cm}\times \mbox{sgn}(q-q') \varphi(q')\, dq'\,dq
\end{eqnarray} 
where $\Sigma$ is the common support of $\varphi(q)$ and $\phi(q)$. We point out that our ability to arrive at the above expression has been made possible by extending the formulation in a rigged Hilbert space.

If $\opr{H}^{\times}$ and $\mcal{T}$ are to be canonically conjugate in the distributional sense, then the second term must identically vanish for all $\varphi(q), \phi(q)\in\Phi(\Re)$, while the first term must identically reduce to $i\hbar\,\dirac{\phi}{\varphi}$. Because $\varphi$ and $\phi$ are arbitrary and $\mbox{sgn}(q-q)$ is not identically zero, the former is satisfied if and only if $T(q,q')$ satisfies the partial differential equation
\begin{equation} \label{kerneldiff}
 -\frac{\hbar^2}{2\mu}\frac{\partial^2 T(q,q')}{\partial q^2}+\frac{\hbar^2}{2\mu} \frac{\partial^2 T(q,q')}{\partial {q'}^2}+ \left(V(q)-V(q')\right)T(q,q') =0.
\end{equation}
On the other hand the later condition is satisfied if and only if $T(q,q')$ satisfies the boundary condition
\begin{equation}\label{origbound}
	\frac{d T(q,q)}{dq} + \left.\frac{\partial T(q,q')}{\partial q}\right|_{q=q'} + 	\left.\frac{\partial T(q,q')}{\partial q'}\right|_{q'=q}=1.
\end{equation}
for all $q,q'\in\Re$. The boundary condition (\ref{origbound}) defines a family of operators canonically conjugate to the extended Hamiltonian in the sense required by equation (\ref{condition}). This is a reflection of the fact that there are numerous operators that are canonically conjugate to a given Hamiltonian. 

The immediate problem now is how to fix the boundary condition on $T(q,q')$ such that (\ref{kerneldiff}) yields a solution satisfying the quantum-classical-correspondence boundary condition (\ref{transition}), and at the same time satisfying the boundary condition (\ref{origbound}). Moreover, it is appropriate to require that the solution to (\ref{kerneldiff}) is unique. Again we appeal to our transfer principle. We find the set of boundary conditions satisfied by the free particle kernel that ensures that the corresponding solution to the time kernel equation is unique for the free particle. For this case the time kernel equation reduces to
\begin{equation}\label{free123}
 -\frac{\hbar^2}{2\mu}\frac{\partial^2 T(q,q')}{\partial q^2}+\frac{\hbar^2}{2\mu} \frac{\partial^2 T(q,q')}{\partial {q'}^2} =0.
\end{equation}
The general solution to this equation is
\begin{equation}\label{gamama}
T(q,q')=f(q+q')+g(q-q').
\end{equation}
For the free particle, we already have $T(q,q')=\frac{1}{4}(q+q')$. Now we have to identify the set of boundary conditions that isolates this known solution from the general solution. 

By inspection $T(q,q')$ satisfies both (\ref{kerneldiff}) and (\ref{origbound}), and it satisfies the conditions
\begin{equation}\label{equiv}
T(q,q)=\frac{q}{2},\;\;\;\;T(q,-q)=0.
\end{equation}
We now show that these two conditions, when imposed on (\ref{gamama}) uniquely identifies the free particle solution. Note that these conditions impose that $T(q,q')$ is analytic in the neighborhood of the origin. Imposing the second condition of (\ref{equiv}) on (\ref{gamama}) gives $T(q,-q)=f(0)+g(2q)=0$; since this must hold for all $q\in\Re$, it must be that $g(2q)=constant=-f(0)$. On the other hand, imposing the first of (\ref{equiv}) gives $T(q,q)=f(2q)-f(0)=\frac{1}{4}(2q)$. Since $T(q,q')$ satisfies equation (\ref{free123}), $f$ is at least twice continuously differentiable. We can then write $f(2q)=f(0)+f'(0) (2q) + R_2$, where $R_2$ is the remainder in the expansion. Thus $T(q,q)=f'(0) (2q)+R_2=\frac{1}{4} (2q)$, which implies that $f'(0)=\frac{1}{4}$ and $R_2=0$. This finally implies that $T(q,q')=\frac{1}{4}(q+q')$, reproducing the solution we know for the free particle.

By our assumed transfer principle, we impose the same boundary conditions (\ref{equiv}) on the solution to the time kernel equation (\ref{kerneldiff}). That the boundary conditions (\ref{equiv}) guarantee that boundary condition (\ref{origbound}) is satisfied and that they impose symmetry on $T(q,q')$ under the interchange of its arguments will be shown below for entire analytic potentials. 

We claim that equations (\ref{kerneldiff}) and (\ref{equiv}) constitute the supraquantization of the local time of arrival consistent with the correspondence principle. We will explicitly demonstrate this claim in Section-\ref{examples} for the harmonic and anharmonic oscillators, and separately demonstrate in Section-\ref{entireanalytic} for entire analytic potentials. 

\subsection{Canonical Form of the Time Kernel Equation}\label{molala}

In what to follow, we will find it convenient to prove our above assertion by solving the time kernel equation (\ref{kerneldiff}) in canonical form. This is accomplished  by performing a change in variable from $(q,q')$ to  $(u=q+q',v=q-q')$. The differential equation (\ref{kerneldiff}) and its accompanying boundary condition (\ref{equiv}) then assume the form
\begin{equation}\label{newdequation}
		-2\frac{\hbar^2}{\mu}\frac{\partial^2 T}{\partial u\,\,\partial v} + 				\left(V\left(\frac{u+v}{2}\right)-V\left(\frac{u-v}{2}\right)\right) T(u,v)=0,
\end{equation}
\begin{equation}\label{newboundary}
			T(u,0)=\frac{u}{4},\;\;\;T(0,v)=0.	
\end{equation}
The boundary conditions (\ref{newboundary}) impose that the solution to equation (\ref{newdequation}) is analytic in $u$ and $v$. In solving for equation (\ref{newdequation}), we will then seek an analytic solution in powers of $u$ and $v$,
\begin{equation}\label{jojo}
 	T(u,v)=\sum_{m,n\geq 0}\alpha_{m,n} \,u^m v^n
\end{equation}
 where the $\alpha_{m,n}$'s are constants determined by equations (\ref{newdequation}) and (\ref{newboundary}) for a given potential. 
 
Assuming a solution of the form (\ref{jojo}), translates the boundary conditions (\ref{newboundary}) to the boundary condition on the expansion coefficients $\alpha_{m,n}$:
\begin{equation}\label{bobo}
	\alpha_{m,0}=\frac{1}{4}\delta_{m,1} \;\;\;
	\alpha_{0,n}=0
\end{equation}
for all $m$ and $n$. We arrive at them as follows. The first boundary condition gives $T(u,0)=\sum_m \alpha_{m 0}\, u^m=\frac{1}{4}u$, which implies the first of equations (\ref{bobo}). And the second boundary condition gives $T(0,v)=\sum_n \alpha_{0 n}\,v^n=0$, which implies the second of equation (\ref{bobo}). However, in our proof below, we will find it convenient to extend the summation in (\ref{jojo}) to negative values of $m$ and $n$; the analicity of the solution is then imposed by adjoining to (\ref{bobo}) the condition that $\alpha_{m,n}=0$ when either $m$ or $n$ is negative or when both are negative.

To show the uniqueness of the solution to (\ref{newdequation}) for a given potential, we will assume the existence of two distinct solutions, say, $T_1 (u,v)$ and $T_2 (u,v)$. Then the function $S(u,v)=T_1 (u,v)-T_2 (u,v)$ satisfies the time kernel equation. $S(u,v)$ then satisfies the boundary conditions $S(u,0)=0$ and $S(0,v)=0$. Since $T_1 (u,v)$ and $T_2 (u,v)$ are both analytic, $S(u,v)$ must itself be analytic. Then $S(u,v)$ can be expanded in $u$ and $v$ in the same way that $T_1$ and $T_2$ can be expanded,
\begin{equation}\label{bulaklak}
	S(u,v)=\sum_{m,n\geq 0}\eta_{m,n}\, u^m\, v^n\, .
\end{equation}
Now the boundary condition satisfied by the expansion coefficients are $\eta_{m,0}=0$ and $\eta_{0,n}=0$ for all $m$ and $n$.  The solution is unique if all the expansion coefficients are identically zero or $S(u,v)$ identically vanishes. We will show below that the solutions for entire analytic potentials are unique.

Now we can address the concern raised earlier whether the assumed properties of $T(q,q')$ are sufficient to ensure that the original boundary condition (\ref{origbound}) is satisfied.  With the assumed form of the solution (\ref{jojo}), the solution in the original coordinates will be in the form
\begin{equation}
T(q,q')=\sum_{m\geq 1, n\geq 0} \alpha_{mn} (q+q')^m \, (q-q')^n.
\end{equation}
Evaluating this at $q=q'$, we have $T(q,q)=2\alpha_{1,0} q$, and arrive at $T'(q,q)=\frac{1}{2}$, because of the boundary condition $\alpha_{1,0}=\frac{1}{4}$. On the other hand we arrive at the following
\begin{eqnarray}
\left.\frac{\partial T(q,q')}{\partial q}\right|_{q=q'}&=&\alpha_{1,0}+\sum_{m\geq 1}\alpha_{m,1}\, 2 q',\nonumber\\
\left.\frac{\partial T(q,q')}{\partial q'}\right|_{q'=q}&=&\alpha_{1,0}-\sum_{m\geq 1}\alpha_{m,1}\, 2 q \nonumber.
\end{eqnarray}
However, $T(q,q')$ is symmetric, i.e. $T(q,q')=T(q',q)$, so that $\alpha_{m,n}=0$ for odd $n$. The second terms of the above equations then vanish and they only take contribution from the first terms. With $\alpha_{1,0}=\frac{1}{4}$, the boundary condition (\ref{origbound}) is then satisfied. We note that we have appealed to the assumed symmetry of $T(q,q')$, but this is not totally necessary, because, as what will be shown below, the boundary conditions (\ref{equiv}) are sufficient to impose the symmetry of $T(q,q')$.

\section{Explicit Examples}\label{examples}

Before we prove our above assertion, we will explicitly demonstrate in this section our claim for two specific systems: the harmonic and the anharmonic oscillators. We will first solve for the local time of arrival in the neighborhood of the origin using
\begin{equation}
t_0(q,p)=\sum_{k=0}^{\infty} (-1)^k \,T_k (q,p)
\end{equation}
where the iterates $T_k$'s are generated through the following recurrence relation,
\begin{equation}\label{x0x}
T_0(q,p)=-\mu \frac{q}{p},
\end{equation}
\begin{equation}\label{yr0}
T_k (q,p)= -\frac{\mu}{p}\int_0^q \frac{\partial V}{\partial q'} \frac{\partial T_{k-1}}{\partial p}\, dq',
\end{equation}
obtained from the general expressions (\ref{inini}) and (\ref{mowcow}) by setting $x=0$.
 
We will then compare the local time of arrival with the Wigner-Weyl transform of the time kernel,
\begin{equation}\label{transkerker}
\mathcal{T}_{\hbar}(q,p)=\!\! \int_{-\infty}^{\infty}\!\!\!\transker\,\exp\left(-i\frac{v\,p}{\hbar}\right) \,dv.
\end{equation}
($\mathcal{T}_{\hbar}(q,p)$ is real valued and odd with respect to $p$.) For the harmonic oscillator we will find that the local time of arrival and $\mcal{T}_{\hbar}$ coincide; and for the anharmonic oscillator it is only in the limit of vanishing or infinitesimal $\hbar$ that $\mcal{T}_\hbar$ reproduces the local time of arrival at the origin. 

\subsection{The Harmonic Oscillator}
\subsubsection{Global and Local Time of Arrivals}

The potential for the harmonic oscillator is $V(q)=\frac{1}{2}\mu \omega^2\,q^2$. Substituting the potential back into the general expression for the global time of arrival (\ref{classpass}) yields,
\begin{equation}
T_0(q,p)=-\frac{1}{\omega}\tan^{-1}\left(\frac{\mu\omega\,q}{p}\right).
\end{equation}
We will show below that this can be derived via the local time of arrival.

Substituting the potential in equation (\ref{yr0}), we generate the first two iterates of the local time of arrival,
\begin{equation}
T_1=-\frac{1}{3}\mu^3 \omega^2 \frac{q^3}{p^3}\nonumber,\;\;\;
T_2=-\frac{1}{5} \mu^5 \omega^4 \frac{q^5}{p^5}\nonumber.
\end{equation}
From these iterates, we infer that for every $k$, the $k$-th iterate is given by
\begin{equation}\label{o123}
T_k=-\alpha_k \mu^{2k+1}\omega^{2k} \frac{q^{2k+1}}{p^{2k+1}},
\end{equation}
where the $\alpha_k$'s are constants to be determined. These constants are  determined as follows. We shift index $k\rightarrow (k-1)$ in $T_k$ to get the expression for $\alpha_{k-1}$. We then substitute $T_{k-1}$ and the potential back in the right hand side  of equation (\ref{yr0}) to yield
\begin{equation}\label{young}
-\frac{\mu}{p}\int_0^q \frac{\partial V}{\partial q'}\frac{\partial T_{k-1}}{\partial p}\, dq'=-\alpha_{k-1} \frac{(2k-1)}{(2k+1)}\mu^{2k+1} \omega^{2k} \frac{q^{2k+1}}{p^{2k+1}}
\end{equation}
If expression (\ref{o123}) holds for all $k$, then the right hand sides of equations (\ref{o123}) and (\ref{young}) must be equal for all $k$. Strict equality is then satisfied if and only if the $\alpha_k$'s satisfy the following recurrence relation among themselves,
\begin{equation}
\alpha_{k}=\frac{(2k-1)}{(2k+1)}\alpha_{k-1},
\end{equation}
subject to the initial value $\alpha_0=1$. This can in turn be solved to give  $\alpha_{k}=(2k+1)^{-1}$. The local time of arrival is then given by
\begin{equation}\label{mumu}
t_0(q,p)=-\sum_{k=0}^{\infty}\frac{(-1)^k}{2k+1}\mu^{2k+1}\omega^{2k}\frac{q^{2k+1}}{p^{2k+1}}.
\end{equation}
$t_0(q,p)$ can be summed within its region of convergence in the phase space, and the result coincides with the global one in the same region.

In the following we will show that $\mcal{T}_{\hbar}(q,p)=t_0(q,p)$, and this is just a special case of our general result on the equality of $\mcal{T}_{\hbar}(q,p)$ and $t_0(q,p)$ for linear systems, i.e. systems with linear classical equations of motion.
 
\subsubsection{Supraquantization of the Local Time of Arrival}
Substituting the potential in equation (\ref{newdequation}) gives the corresponding time kernel equation to solve for the harmonic oscillator,
\begin{equation}\label{5harmonic}
		-2\frac{\hbar^2}{\mu}\,\frac{\partial^2 T}{\partial u \partial v}(u,v) + 
										\frac{\mu \omega^2}{2}\,uv\, T(u,v) =0,
\end{equation}
subject to the boundary conditions (\ref{equiv}). We assume a solution of the form
\begin{equation*}\label{assharm}
	T(u,v)=\sum_{m,n} \alpha_{m,n}\, u^m\, v^n
\end{equation*}
where the $\alpha$'s are constants to be determined, subject to the boundary conditions $\alpha_{m,0}=\frac{1}{4} \delta_{m,1}$, $\alpha_{0,m}=0$ for all $m$, and $\alpha_{m,n}=0$ for $m,n < 0$. Substituting the assumed solution back into equation (\ref{5harmonic}), we arrive at
\begin{equation*}
		-2\frac{\hbar^2}{\mu} \sum_{m,n} \alpha_{m,n}\, mn\, u^{m-1}\, v^{n-1}
				+ \frac{\mu \omega^2}{2} \sum_{m,n} \alpha_{m,n}\, u^{m+1}\, v^{n+1} =0.
\end{equation*}
Shifting indices in the second term, $m\rightarrow (m-1)$ and $n\rightarrow (n-1)$, and collecting like terms, we get
\begin{equation*}
		\sum_{m,n}\left(-2\frac{\hbar^2}{\mu} mn\, \alpha_{m,n} + 	
			\frac{\mu \omega^2}{2}\, \alpha_{m-2,n-2}\right)\, u^{m-1}\, v^{n-1} = 0.
\end{equation*}
Since $u$ and $v$ are arbitrary, the quantity in the bracket must vanish for all values of $u$ and $v$, dictating the coefficients to satisfy the recurrence relation
\begin{equation}\label{5recurharm}
		\alpha_{m,n}=\left(\frac{\mu^2 \omega^2}{4 \hbar^2}\right)\frac{1}{m\cdot n}\,
								\alpha_{m-2,n-2}.
\end{equation}
Solving the time kernel then reduces to solving this recurrence relation among the coefficients of the assumed solution.

First for odd $n$. Since $\alpha_{m,n}=0$ for negative $n$, let us start from $n=1$. For $n=1$ we get the proportionality $\alpha_{m,2}\propto \alpha_{m-1,-1}$; but the coefficients vanish for negative $n$ for all $m$; thus the right hand side of the proportionality vanishes and consequently $\alpha_{m,1}=0$ for all $m$. Now if for some fixed odd $n'$, $\alpha_{m,n'}=0$ for all $m$, then for the next odd $n'+2$, $\alpha_{m,n'+2}\propto \alpha_{m-1,n'}=0$ for all $m$. Since we have already shown that $\alpha_{m,1}=0$ for all $m$, it follows that $\alpha_{m,n}=0$ for all odd $n$, for all $m$. The odd powers of $v$ then do not contribute. 

We remark that the vanishing of the contributions for odd $n$ is significant. We recall that the solution $T(q,q')$ to the time kernel equation in the original form must satisfy the boundary condition (\ref{origbound}). And we have earlier noted in Section-\ref{molala} that if $T(q,q')=T(q',q)$ or $T(u,v)=T(u,-v)$, the condition (\ref{origbound}) is automatically satisfied as long as the boundary conditions (\ref{equiv}) are satisfied as well. But the condition $T(u,v)=T(u,-v)$ is equivalent to the vanishing of the odd powers of $v$. Consequently $T(q,q')$ will automatically satisfy (\ref{origbound}). $T(q,q')$ will then satisfy the original boundary condition (\ref{origbound}). This will be shown to be true for the rest of the potentials considered, particularly for entire analytic potentials.

Now for even $n$. Since $\alpha_{m,n}=0$ for negative $n$ and $\alpha_{m,0}$ is specified, we start with $n=2$. For $n=2$ we get the proportionality $\alpha_{m,2}\propto \alpha_{m-2,0}$; but $\alpha_{m',0}\propto \delta_{m',1}$, thus only $m=3$ contribute or $\alpha_{3,2}$ is the only non-vanishing coefficient for $n=2$. For $n=4$ we get $\alpha_{m,4}\propto \alpha_{m-2,2}$, which dictates that only $m=5$ contributes or $\alpha_{5,4}$ is the only non-vanishing coefficient for $n=4$. We see that $n$ and $m$ are not independent from its each other, i.e. they can be index by the same letter, say k. From the first two coefficients we infer that $m=2k+1$ and $n=2k$, $k=1,\, k=2,\, \ldots$. We can prove this by induction. Let for some fixed $k$ that $\alpha_{m=2k+1,n=2k}$ is the only non-vanishing coefficient for $n=2k$. Then for $k+1$, we have $\alpha_{m,2(k+1)}\propto \alpha_{m-2,2k}$; but the only non-vanishing contributions come from $m-2=2k+1$ or $m=2k+3$. Thus for $n=2(k+1)$, only $\alpha_{2k+3,2(k+1)}$ is non-zero. Thus indeed only the coefficients $\alpha_{2k+1,2k}$ are non-vanishing for all $k=1,\,2,\,\dots$. Then the double index recurrence relation (\ref{5recurharm}) reduces to the single index recurrence relation,
\begin{equation}\label{newnew}
			\alpha_k= \left(\frac{\mu^2 \omega^2}{4 \hbar^2}\right)
						\frac{1}{(2k+1)2k}\; \alpha_{k-1},
\end{equation}
subject to the initial value $\alpha_0=\alpha_{1,0}=\frac{1}{4}$. The solution to equation (\ref{newnew}) is
\begin{equation}
		\alpha_k=\frac{1}{4}\left(\frac{\mu \omega}{2 \hbar}\right)^2\frac{1}{(2k+1)!}.
\end{equation}

Substituting the non-vanishing coefficients back in the assumed solution yields the solution to time kernel equation for the harmonic oscillator,
\begin{eqnarray}
		T(u,v)=\frac{\hbar}{2 \mu \omega} \sum_{k=0}^{\infty} \frac{1}{(2k+1)!}
				\left(\frac{\mu \omega}{2 \hbar}\right)^{2k+1}\; u^{2k+1}\, v^{2k}\label{uhog}.
\end{eqnarray}
Evidently $T(u,v)$ converges everywhere in the $uv$-plane. Moreover, the solution (\ref{uhog}) is unique. This follows from the fact that $S(u,v)$ (see equation (\ref{bulaklak})) satisfies the time kernel equation, and it satisfies the boundary conditions $\eta_{0,n}=0$, $\eta_{m,0}=0$ for all $m$, $n$ on its coefficients. The recurrence relation on the coefficients $\eta_{m,n}$ will be the same as those of the $\alpha_{m,n}$'s. Since the non-vanishing contributions in $T(u,v)$ come only from the boundary condition $\alpha_{m,0}=\frac{1}{4}\delta_{m,1}$, $S(u,v)$ will be identically zero because $\eta_{m,0}=0$ for all $m$. The analytic solution $T(u,v)$ is then unique. This observation holds for the rest of the potentials considered here.

Transforming back to $(q,q')$ and substituting $T(q,q')$ back into equation (\ref{timekernel}) yields the time kernel for the harmonic oscillator, 
\begin{equation}
	\kernel{\mcal{T}}=\frac{1}{2 i\, \omega}\mbox{sgn}(q-q') \sum_{k=0}^{\infty} \frac{1}{(2k+1)!}
				\left(\frac{\mu \omega}{2 \hbar}\right)^{2k+1}\; (q+q')^{2k+1}\, (q-q')^{2k}.
\end{equation}
This likewise converges everywhere in the $qq'$-plane. Because $T(q,q')$ is everywhere absolutely convergent, $\mcal{T}$ is a generalized observable in $\Phi$ (see Appendix). Now we can finally show that the generalized time of arrival operator reduces to the local time of arrival in the classical limit, as prescribed by equation (\ref{transition}). Using the identity \cite{gel2}
\begin{equation}\label{identity}
\int_{-\infty}^{\infty}\sigma^{m-1}\mbox{sgn}(\sigma)\,\exp(-ix\sigma)\,d\sigma= \frac{2(m-1)!} {i^m} x^{-m},
\end{equation}
we can perform the indicated transformation to give
\begin{eqnarray*}
		\mathcal{T}_{\hbar}(q,p)&=&\!\! \int_{-\infty}^{\infty}\!\!\!
			\transker\,\exp\left(-i\frac{v\,p}{\hbar}\right) \,dv \nonumber \\
			&=&\frac{1}{2 i\, \omega} \sum_{k=0}^{\infty} \frac{1}{(2k+1)!}
				\left(\frac{\mu \omega}{2 \hbar}\right)^{2k+1}\; (2q)^{2k+1} \int_{-\infty}^{\infty}
		v^{2k} \mbox{sgn}(v) \,\exp\left(-i\frac{v\,p}{\hbar}\right) \,dv 
				\nonumber \\
			&=&-\frac{1}{\omega}\sum_{k=0}^{\infty} \frac{(-1)^k}{2k+1}
					\left(\frac{\mu\omega q}{p}\right)^{2k+1}.
\end{eqnarray*}
We find that $\mcal{T}_{\hbar}$ coincides exactly with the local time of arrival in the neighborhood of the origin for the harmonic oscillator given by equation (\ref{mumu}).

\subsection{The Anharmonic Oscillator}
\subsubsection{Global and Local Time of Arrivals}

In the previous section the Weyl-Wigner transform of the time kernel exactly reproduces the local time of arrival at the origin. But for non-linear systems, systems with non-linear equations of motions, we demonstrate that only in the limit of vanishing or infinitesimal $\hbar$ that the local time of arrival (at the origin) is recovered. Let us consider the anharmonic oscillator with the potential $V=\lambda q^4$. The global time of arrival is symbolically given by
\begin{equation}
T_0(q,p)=-\mbox{sgn}(p)\sqrt{\frac{\mu}{2}} \int\limits_0^q\!\!\! \frac{dq'}{\sqrt{H(q,p) -\lambda {q'}^4}}.
\end{equation}
The above expression can be integrated explicitly, but its exact closed integral is not important to our present purposes.

What is important to us is the local time of arrival in the neighborhood of the origin. Following the same procedure we have employed above in determining for the iterates of the harmonic oscillator, we find that the $k$-th iterate of the local time of arrival is given by
\begin{equation}\label{gogo}
T_k=\frac{1}{4}\,{\frac {{2}^{k}\sqrt {\pi }\Gamma (-k-\frac{1}{4})}{\Gamma (\frac{3}{4})\Gamma (
-k+\frac{1}{2})}}\alpha_k \mu^{k+1}\lambda^k \frac{q^{4k+1}}{p^{2k+1}}.
\end{equation}
One can prove this by induction using equation (\ref{yr0}). Substituting $T_{k}$ back in equation (\ref{series}) yields the local time of arrival at the origin,
\begin{equation}\label{noo}
t_0(q,p)=\frac{1}{4}\frac{\sqrt{\pi}}{\Gamma\left(\frac{3}{4}\right)} \sum_{k=0}^{\infty} \frac{(-2)^{k} \Gamma\left(-k-\frac{1}{4}\right)}{\Gamma\left(\frac{1}{2}-k\right)}\, \mu^{k+1}\lambda^{k} \frac{q^{4k+1}}{p^{2k+1}}.
\end{equation}

In the following we will show that $\mcal{T}_{\hbar}(q,p)=t_0(q,p)+\order{2}$. And this is just a special case of our general result on the equality of $\mcal{T}_{\hbar}(q,p)$ and $t_0(q,p)$ only in the limit of vanishing or infinitesimal $\hbar$ for non-linear systems, i.e. systems with non-linear classical equations of motion.

\subsubsection{Supraquantization of the Local Time of Arrival}
Substituting the potential equation in equation (\ref{newdequation}) gives the corresponding time kernel equation for the anharmonic oscillator,
\begin{equation}\label{anharmpde}
			-2 \frac{\hbar^2}{\mu} \frac{\partial^2 T}{\partial v \, \partial u} 
						+ \frac{\lambda}{2} \left(u^3v + u v^3\right)T(u,v)=0,
\end{equation}
subject to the same boundary conditions. Again we assume the most general form of the solution to the time kernel equation (\ref{anharmpde}), 
\begin{equation*}
		T(u,v)=\sum_{m,n} \alpha_{m,n} u^m \, v^n\,
\end{equation*}
where the $\alpha_{m,n}$'s satisfy the boundary conditions $\alpha_{m,0}=\frac{1}{4}\delta_{m,1}$ and $\alpha_{0,n}=0$ for all $m$ and $n$, and $\alpha_{m,n}=0$ for negative $m$ and $n$ to identify the particular solution we seek. 

Substituting the assumed solution back to equation (\ref{anharmpde}) gives the following recurrence relation for the coefficients,
\begin{equation}\label{5recuranharm}
		\alpha_{m,n} = \left(\frac{\mu \lambda}{4 \hbar^2}\right) 
				\frac{1}{mn} \left(\alpha_{m-4,n-2}+ \alpha_{m-2,n-4}\right).
\end{equation}
First let us consider the coefficients for odd powers of $v$ or for odd $n$. Since the coefficients vanish for negative $m$ and $n$, we start from $n=1$. For $n=1$ we get the proportionality $\alpha_{m,1}\propto (\alpha_{m-4,-1}+\alpha_{m-2,-3})$. But $\alpha_{m,n}=0$ for $n<0$ for all $m$, so that $\alpha_{m,1}=0$ for all $m$. For $n=3$ we get the proportionality $\alpha_{m,3}\propto (\alpha_{m-4,1}+\alpha_{m-2,-1})$.   Since $\alpha_{m,1}=0$ for all $m$ and $\alpha_{m,n}=0$ for $n<0$, $\alpha_{m,3}=0$ for all $m$ as well. Now if for some odd $n$, $\alpha_{m,n}=0$ for all $m$, it follows from (\ref{5recuranharm}) that for the next odd number $n+2$, $\alpha_{m,n+2}=0$ for all $m$. Thus odd powers of $v$ vanish. This assures us that $T(q,q')$ satisfies the boundary condition (\ref{origbound}).

Let us now consider the even powers of $v$. For $n=2$ we get the proportionality $\alpha_{m,2}\propto (\alpha_{m-4,0}+\alpha_{m-2,-2})$. Only $m=5$ contributes because $\alpha_{m,n}=0$ for negative $n$ and $\alpha_{m,0}=\frac{1}{4} \delta_{m,1}$; thus for $n=2$ only $\alpha_{5,2}$ contributes. For $n=4$ we get the proportionality $\alpha_{m,4}\propto (\alpha_{m-4,2}+\alpha_{m-2,0})$. There are only two contributions: $m=3$, corresponding to $\alpha_{1,0}$, and $m=9$, corresponding to $\alpha_{5,2}$; thus for $n=4$ only $\alpha_{3,2}$ and $\alpha_{9,2}$ contribute. Continuing in this manner, we arrive at the following first few sequences of nonvanishing contributions,
\begin{eqnarray*}
n=0: & & \alpha_{1,0} \\
n=2: & & \alpha_{5,2} \\
n=4: & & \alpha_{9,4},\;\;\alpha_{3,4}\\
n=6: & & \alpha_{13,6},\;\;\alpha_{7,6}\\
n=8: & & \alpha_{17,8},\;\;\alpha_{11,8},\;\;\alpha_{5,8}\\
n=10: & & \alpha_{21,10},\;\;\alpha_{15,10},\;\;\alpha_{9,10}
\end{eqnarray*}
By inspection the non-vanishing coefficients can be grouped in two groups. Let $n=2j$ for $j=0,1,2,\dots$. The contributing coefficients can then be written in the form $\alpha_{m(j),2j}$, where for 
\begin{eqnarray*}
j&\!=\!&\mbox{odd},\;\; m(j)=\left(j+4\right),\, \left(j + 4\right) +6,\,\left(j+4\right)+12\,\dots, 2j+1,\\ 
j&\!=\!&\mbox{even},\;\; m(j)=\left(j+1\right),\, \left(j+1\right)+6,\, \left(j+1\right)+12\,\dots,\,2j+1.
\end{eqnarray*}
Evidently for a given $j$ there are $\{j\}$ contributing $m$'s, in which $\{j\}=\frac{1}{2}(j-1)+1$ for $j$=odd, and $\{j\}=(\frac{1}{2}j-1)+1 $ for $j=$even. This can be proved by induction.

With the arrangement above for the coefficients, we can sum along the vertical.
The above results suggest that the solution can be written in the following form
\begin{equation}\label{secondform}
		T(u,v)=\frac{1}{4}\sum_{k=0}^{\infty} \sum_{j=2k}^{\infty} 
					\beta_{k,j} \, u^{4j+1-6k}\, v^{2j}
\end{equation}
where $\beta_{k,j}$'s are proportional to the non-vanishing coefficients, i.e. $\beta_{k,j}=4\,\alpha_{4j+1-6k,2j}$, $\beta_{0,0}=1$, for $k\geq 0$ and $j\geq 2k$. 
Substituting equation (\ref{secondform}) back into equation (\ref{anharmpde}), we get the following recurrence relation for the $\beta_{k,j}$'s,
\begin{equation}\label{newrecur}
			\beta_{k,j}=\left(\frac{\mu \lambda}{4 \hbar^2}\right) 
					\frac{1}{(4j+1-6k)\cdot 2j} \left(\beta_{k,j-1}+\beta_{k-1,j-2}\right).
\end{equation}
This recurrence relation (\ref{newrecur}) holds for all values of $k$ and $j$ restricted in the assumed solution as long as we agree to set $\beta_{k,j}=0$ when both or either of $k$ and $j$ is negative, or when $j<2k$. 

First let us solve for $\beta_{0,j}$ for all $j\geq 1$ given $\beta_{0,0}=1$. Setting $k=0$ in equation (\ref{newrecur}) we arrive at the recurrence relation
\begin{equation}\label{yoyo}
			\beta_{0,j}=\left(\frac{\mu\lambda}{4\hbar^2}\right)
					\frac{1}{(4j+1)\cdot 2j} \beta_{0,j-1},
\end{equation}
the $\beta_{k,j}$'s being zero for $k<0$. Let us define 
\begin{eqnarray}
\lambda_r^{(j,k)}=\prod_{l=0}^{r}\frac{1}{(4(j-l)+1-6k)},\label{beng}
\end{eqnarray}
where the value $r=-1$ is allowed. Equation (\ref{yoyo}) can be solved recursively to give
\begin{equation}\label{solution0}
			\beta_{0,j}	=\frac{1}{j!}\left(\frac{\mu\lambda}{8\hbar^2}\right)^j 
				\frac{1}{(-4)^{j+1}}{\frac {\Gamma (-j-\frac{1}{4})}{\Gamma (\frac{3}{4})}}
\end{equation}
valid for all $j\geq 0$. Given $\beta_{0,j}$ we can proceed in determining of the coefficients.

To solve for $\beta_{k,j}$ for arbitrary $k$ and $j$, we assume that we know the solution for $(j-1)$ for all $j$ in equation (\ref{newrecur}). This reduces the problem of solving the recurrence relation for some fix $j$. We shift index $k\rightarrow (k-1)$ in equation (\ref{newrecur}) and substitute it back to equation (\ref{newrecur}). We do this repeatedly until we arrive at the following result,
\begin{equation}\label{tofurthersolve}
	\beta_{k,j}=\sum_{r=0}^{j-2k} \left(\frac{\mu\lambda}{4\hbar^2}\right)^{r+1}  						\frac{\beta_{k-1,j-2-r}}{2^{r+1}} \frac{(j-1-r)!}{j!}\lambda_r^{(j,k)}.
\end{equation}
This can be proven by induction. So for $k=1$ equation (\ref{tofurthersolve}) yields \begin{eqnarray}\label{kiko}
\beta_{1,j}&=&\frac{1}{j!}\left(\frac{\mu\lambda}{8\hbar^2}\right)^{j-1} \, \sum_{r=0}^{j-2} (j-r-1) \, \lambda_{j-r-3}^{(j-2-r,0)}\, \lambda_{r}^{(j,1)}\nonumber\\
&=& \frac{1}{j!}\left(\frac{\mu\lambda}{8\hbar^2}\right)^{j-1}\, \left[ \frac{1}{2}\,{\frac {\Gamma (\frac{3}{4})\left (-1\right )^{j}\sqrt {2}\Gamma (-j+\frac{5}{4})
}{{4}^{j}\pi }}\right.\nonumber\\
& & +\left.{\frac {1}{192}}\,{\frac {\left (4\,j-5\right )\left (
2\,j+1\right )\left (-1\right )^{j-2}\Gamma (-j+\frac{7}{4})\Gamma (-j+\frac{5}{4})}{{
4}^{j-2}\Gamma (\frac{3}{4})\Gamma (\frac{9}{4}-j)}}\right].
\end{eqnarray}
The rest of the contributing coefficients for other $k$'s can be determined similarly.

However, the explicit forms for $\beta_{0,j}$ and $\beta_{1,j}$ suggest a simplification for $\beta_{k,j}$. These coefficients can be explicitly written in the form,
\begin{equation}\label{google}
\beta_{k,j}=\frac{1}{j!}\left(\frac{\mu\lambda}{8\hbar^2}\right)^{j-k} \rho_{k,j}
\end{equation}
for some constants $\rho_{k,j}$. These constants are found by substituting (\ref{google}) back in both sides of equation (\ref{tofurthersolve}), with the appropriate shifting of indices in the right hand side. Doing so leads to the following recurrence relation
\begin{equation}\label{jojof}
\rho_{k,j}=\sum_{r=0}^{j-2k}(j-r-1)\, \lambda_r^{(j,k)}\, \rho_{k-1,j-2-r}
\end{equation}
for all $k\geq 1$ and $j\geq 2k$. The initial value that determines all the constants $\rho_{k,j}$ is defined by (\ref{solution0}). Comparing equation (\ref{solution0}) and equation (\ref{google}) for $k=0$ gives
\begin{eqnarray}
	\rho_{0,j}&=&\frac{1}{(-4)^{j+1}}{\frac {\Gamma (-j-\frac{1}{4})}{\Gamma (\frac{3}{4})}},
\end{eqnarray}
valid for all $j\geq0$. Equation (\ref{jojof}) can be solved explicitly given the initial value. We don't need to write its explicit solution. 

Substituting $\beta_{k,j}$ back in equation (\ref{secondform}), the solution to the time kernel equation for the anharmonic oscillator assumes the form
\begin{equation}\label{solsol}
	T(u,v)=\frac{1}{4}\sum_{k=0}^{\infty} \sum_{j=2k}^{\infty} \frac{\rho_{k,j}}{j!} 					\left(\frac{\mu\lambda}{8\hbar^2}\right)^{j-k} \, u^{4j+1-6k} \, v^{2j}.
\end{equation}
In arriving at this solution we have assumed above that the solution absolutely converges in the neighborhood of the origin in the $uv$-plane, so that the contributing terms can be rearranged at will. However, we can only assert at this moment absolute convergence in the entire $uv$-plane of the first two leading terms in the solution. Equation (\ref{solsol}) can be written in the form
\begin{equation}
T(u,v)=T_0(u,v)+T_1(u,v)+\dots,
\end{equation}
where the subscripts denote the corresponding term for a given $k$. For $k=0$ and $k=1$, we have the following explicit closed forms,
\begin{eqnarray}
T_0(u,v)&=&\frac{1}{4}u\,\, _{0}F_1\!\left(\frac{5}{4};\frac{\mu\lambda}{32\hbar^2}u^4 v^2\right)\\
\nonumber\\
T_1(u,v)&=&-\frac{1}{96}\left(\frac{\mu\lambda}{8\hbar^2}\right)u^3v^4\;_{1}F_{2}\left(1; 3, \frac{7}{4};\frac{\mu\lambda}{32\hbar^2}u^4 v^2\right)\nonumber\\
& &+\frac{5}{96}\left(\frac{\mu\lambda}{8\hbar^2}\right)u^3v^4\;_{1}F_{2}\left(1; 3, \frac{5}{4};\frac{\mu\lambda}{32\hbar^2}u^4 v^2\right)\nonumber\\
& &+\frac{1}{720}\left(\frac{\mu\lambda}{8\hbar^2}\right)^2u^7v^6\;_{1}F_{2}\left(2; 4, \frac{9}{4};\frac{\mu\lambda}{32\hbar^2}u^4 v^2\right),
\end{eqnarray}
where $_{p}F_{q}$ is the generalized hypergeometric function. $_{p}F_{q}$ converges everywhere if $p\leq q$. The first two leading terms then converge everywhere in the $uv$-plane. 

Transforming back to $(q,q')$, the time kernel for the anharmonic oscillator assumes the form
\begin{equation}
	\kernel{\mcal{T}}=\frac{\mu}{4i\hbar}\mbox{sgn}(q-q')\sum_{k=0}^{\infty} 			\sum_{j=2m}^{\infty} 
			\frac{\rho_{k,j}}{j!} \left(\frac{\mu\lambda}{8\hbar^2}\right)^{j-k} \,
 			(q+q')^{4j+1-6k} \, (q-q')^{2j}.
\end{equation}
The first two groups of terms, corresponding to $k=0$ and $k=1$, converge everywhere in the $qq'$-plane so that they are functionals in the $\Phi^{\times}$ and define a generalized operator in $\Phi$. We have not been able to make a conclusion on the convergence of the rest of the group of terms. We note though that, since $\kernel{\mcal{T}}$ must be a functional in $\Phi^{\times}$, it is not necessary that the sum converges in the usual sense; it is sufficient that it converges in the distributional sense in $\Phi^{\times}$ (see Gel'fand \& Shilov (1964) page-368 for a discussion on the convergence of functionals in $\Phi$).  Performing the indicated transformation, we have
\begin{eqnarray*}
		\mathcal{T}_{\hbar}(q,p)&=&\!\! \int_{-\infty}^{\infty}\!\!\!
			\transker\,\exp\left(-i\frac{v\,p}{\hbar}\right) \,dv \nonumber \\
		  &=&-\frac{\mu}{2} \sum_{k=0}^{\infty}\hbar^{2k}\!\!\! 
				\sum_{j=2k}^{\infty}(-1)^j \,\frac{\rho_{k,j}}{j!}(2j)!  
				\left(\frac{\mu\lambda}{8}\right)^{j-k}\frac{(2q)^{4j+1-6k}}{p^{2j+1}}
				\nonumber\\
		&=&\frac{1}{4}\frac{\sqrt{\pi}}{\Gamma(\frac{3}{4})}\sum_{j=0}^{\infty} 
				\frac{(-2)^{j}\Gamma\left(-\frac{1}{4}-j\right)}
				{\Gamma\left(\frac{1}{2}-j\right)} \mu^{j+1}\lambda^{j} 				\frac{q^{4j+1}}{p^{2j+1}}+\mathcal{O}(\hbar^2)\nonumber\\
		&=&t_0 (q,p) + \mathcal{O}(\hbar^2)
\end{eqnarray*}
where $t_0$ is the local time of arrival of the anharmonic oscillator as given by equation (\ref{noo}). Thus $\mathcal{T}_{\hbar}(q,p)$ reduces to $t_0 (q,p)$ in the limit of vanishing $\hbar$ or infinitesimal $\hbar$. 

\section{Entire Analytic Potentials}\label{entireanalytic}
 In this section we prove that the local time of arrival is completely derivable from the generalized time of arrival operator for systems subject to entire analytic potentials, e.g. potentials in polynomials of $q$. We divide our proof for the linear (systems with linear classical equations of motions) and non-linear (systems with non-linear classical equations of motions) cases. In particular we will show that 
\begin{eqnarray}
\mbox{Linear Systems:}& &\mcal{T}_{\hbar}(q,p)=t_0(q,p)\label{56y}\\
\mbox{Non-Linear Systems:}& &\mcal{T}_{\hbar}(q,p)=t_0(q,p)+\order{2}\label{56x}.
\end{eqnarray}

Our method of proof will not follow the line used in the previous section. It will be sufficient to show that the leading term in $\mcal{T}_{\hbar}(q,p)$, $\mcal{T}_k^0(q,p)$, the term of order $\order{0}$, can be written in the form 
\begin{equation}
\mcal{T}_{\hbar}^0(q,p)=\sum_{k=0}^{\infty} (-1)^k \mcal{T}_k (q,p),
\end{equation}
where the $\mcal{T}_k$'s satisfy the initial value and the recurrence relation
\begin{equation}\label{ini6x}
\mcal{T}_0(q,p)=-\mu \frac{q}{p},
\end{equation}
\begin{equation}\label{recur4xv}
\mcal{T}_k(q,p)=-\frac{\mu}{p} \int_{0}^q \frac{\partial V}{\partial q'} \frac{\partial \mcal{T}_{k-1}}{\partial p}\, dq',
\end{equation}
for all $k$. If they do, then, according to our result in Section-\ref{classtoa}, $\mcal{T}_{\hbar}^0(q,p)$ converges to the local time of arrival in the neighborhood of the origin.

\subsection{Linear Systems}
Now we consider the most general case for linear systems. The most general potential is given by
\begin{equation}\label{linlin}
V(q)=aq+\frac{1}{2}b\,q^2,
\end{equation}
for some constants $a$ and $b$. A constant can be added to $V(q)$, but it does not change the result. Here we prove that for systems subject to the potential (\ref{linlin}) the local time of arrival in the neighborhood of the origin is given by equation (\ref{56y}).

We first solve for the time kernel of the generalized time of arrival operator. Substituting the potential in the time kernel equations lead to solve the following partial differential equation
\begin{equation}\label{kokoko}
-\frac{2\hbar^2}{\mu} \frac{\partial^2 T}{\partial u \partial v} + \left(av + \frac{1}{2} b uv\right) T(u,v)=0,
\end{equation}
subject to the boundary conditions $T(u,0)=\frac{1}{4}u$ and $T(0,v)=0$. We assume the most general solution of the form
\begin{equation}
T(u,v)=\sum_{m,n} \alpha_{m,n} u^m v^n,
\end{equation}
where the coefficients now satisfy the conditions $\alpha_{m,n}=0$ when both or either $m$ and $n$ are negative, and $\alpha_{m,0}=\frac{1}{4}\delta_{m,1}$, $\alpha_{0,n}=0$ for all $n$. Substituting the assumed solution back into equation (\ref{kokoko}) yields the recurrence relation among the coefficients
\begin{equation}\label{lin123}
\alpha_{m,n}=\left(\frac{\mu}{2\hbar^2}\right)\frac{1}{mn}\left(a\, \alpha_{m-1,n-2}+ \frac{1}{2} b\, \alpha_{m-2,n-2}\right).
\end{equation}

First let us show that odd powers of $v$ do not contribute in the solution, i.e. $\alpha_{m,n}=0$ for all odd $n$ for all $m$. For $n=1$ we have the proportionality $\alpha_{m,1}\propto (a\, \alpha_{m-1,-1}+\frac{1}{2}b\, \alpha_{m-2,-1})$. But $\alpha_{m,n}=0$ for all negative $n$, thus $\alpha_{m,1}=0$ for all $m$. Now let us assume that for some fixed odd $n$, $\alpha_{m,n}=0$ for all $m$. Then for the next odd $n'=n+2$, $\alpha_{m,n+2}\propto (a\, \alpha_{m-1,n}+\frac{1}{2}b\, \alpha_{m-2,n})$. But $\alpha_{m,n}=0$ for all $m$, thus $\alpha_{m,n+2}=0$ for all $m$. Since we have shown that $\alpha_{m,1}=0$ for all $m$, it then follows that $\alpha_{m,n}=0$ for all $m$ for all odd $n$. The vanishing of the odd powers of $v$ is significant because it assures us that the boundary condition (\ref{origbound}), necessary to impose canonicality, is satisfied.

Now let us determine the coefficients for even $n$. For $n=2$ we have the proportionality $\alpha_{m,2}\propto (a\, \alpha_{m-1,0}+\frac{1}{2}b\, \alpha_{m-2,0})$. Since $\alpha_{m,0}=\frac{1}{4}\delta_{m,1}$, only $m=2$ contributes in the first term, and $m=3$ for the second term. For $n=4$, $\alpha_{m,4}\propto (a\, \alpha_{m-1,2}+\frac{1}{2}b\, \alpha_{m-2,2})$.  Only $m=3, 4$ contribute for the first term; while $m=4,5$ for the second term. Continuing the process, we arrive at the following first few non-vanishing coefficients,
\begin{eqnarray*}
n=0: & & \alpha_{1,0}\\
n=2: & & \alpha_{2,2},\;\; \alpha_{3,2}\\
n=4: & & \alpha_{3,4},\;\; \alpha_{4,4},\;\; \alpha_{5,4}
\end{eqnarray*}
We find that for every $n=2j$, for integer $j$, there are $j+1$ non-vanishing contributions. In particular, for a given $j$, we can write $\alpha_{m,n}$ in the form $\alpha_{m(j),2j}$, where $m(j)$ takes on one of the following values
\begin{equation*}
m(j)=(j+1),\, (j+2),\, \dots,(2j+1).
\end{equation*}
This can be proven by induction.

The solution to the time kernel equation (\ref{kokoko}) can then be written in the following form
\begin{eqnarray}
T(u,v)&=&\sum_{j=0}^{\infty}\sum_{k=0}^j \alpha_{2j+1-k,2j} \, u^{2j+1-k}\, v^{2j}\nonumber\\
&=& \frac{1}{4} \sum_{k=0}^{\infty}\sum_{j=0}^k \eta_{k,j} \, u^{2j+1-k}\, v^{2j},
\end{eqnarray}
where the $\eta_{k,j}$'s are constants proportional to the non-vanishing $\alpha_{m,n}$'s, with $\eta_{0,0}=1$. Substituting back the second form of the solution to the time kernel equation yields the recurrence relation
\begin{equation}\label{goog}
\eta_{k,j}=\left(\frac{\mu}{2\hbar^2}\right) \frac{\left(a\, \eta_{k-1,j-1}+\frac{1}{2}b\, \eta_{k-1,j}\right)}{2k(2k+1-j)},
\end{equation}
valid for all $k$ and $j$ as long as we set $\eta_{j,k}=0$ for all $0>j>k$. Solving this recursively yields the following form
\begin{equation}\label{upo}
\eta_{k,j}=\left(\frac{\mu}{2\hbar^2}\right)^k\frac{1}{2^k k!} \sigma_{k,j}\, b^{k-j} a^j,
\end{equation}
for some constants $\sigma_{k,j}$. Substituting equation (\ref{upo}) back into (\ref{goog}) gives the recurrence relation for the constants $\sigma_{k,j}$,
\begin{equation}\label{lakay}
\sigma_{k,j}=\frac{1}{(2k+1-j)}\left(\sigma_{k-1,j-1}+\frac{1}{2}\sigma_{k-1,j}\right).
\end{equation}
This can be solved recursively to give
\begin{equation}
\sigma_{k,j}=\sum_{r=1}^{k-j+1}\frac{\sigma_{k-r,j-1}}{2^{r-1}}\prod_{s=0}^{r-1} \frac{1}{(2(k-s)+1-j)}
\end{equation}
with $\sigma_{0,0}=1$. We don't need to evaluate this explicitly. We will just need this recurrence relation in proving equation (\ref{56y}). The solution to the time kernel equation can now be written in the form
\begin{equation}
T(u,v)=\frac{1}{4} \sum_{k=0}^{\infty} \sum_{j=0}^{k} \left(\frac{\mu}{2\hbar^2}\right)^k \frac{1}{2^k k!} \sigma_{k,j}\, b^{k-j} a^j \, u^{2k+1-j}\, v^{2k}.
\end{equation}
It can be shown that $T(u,v)$ is everywhere defined in the $uv$-plane. From the expression for $T(u,v)$, we have
\begin{equation}\label{dung}
T(q,q')=\frac{1}{4}\sum_{k=0}^{\infty}\left(\frac{\mu}{2\hbar^2}\right)^k \frac{1}{2^k k!} (q-q')^{2k} \sum_{j=0}^{k}\sigma_{k,j} b^{k-j} a^j (q+q')^{2k+1-j}
\end{equation}
in the original coordinate.

Equation (\ref{dung}) allows us to finally write the functional of the generalized time of arrival operator in the neighborhood of the origin,
\begin{equation}
\kernel{\mcal{T}}=\frac{\mu}{4i\hbar}\mbox{sgn}(q-q') \sum_{k=0}^{\infty} \sum_{j=0}^{k} \left(\frac{\mu}{2\hbar^2}\right)^k \frac{1}{2^k k!} \sigma_{k,j}\, b^{k-j} a^j \, (q+q')^{2k+1-j}\, (q-q')^{2k}.
\end{equation}
This is a functional for fixed $q$ in $\Phi$. Performing the indicated transformation to go back in the classical regime, we get
\begin{eqnarray}
\mathcal{T}_{\hbar}(q,p)&=&\!\! \int_{-\infty}^{\infty}\!\!\!\transker\,\exp\left(-i\frac{v\,p}{\hbar}\right) \,dv\nonumber\\
&=&\sum_{k=0}^{\infty} \sum_{j=0}^{k} (-1)^{k+1} \mu^{k+1}\frac{(2k)!}{k!}\frac{\sigma_{k,j}}{2^j} b^{k-j} a^j \frac{q^{2k+1-j}}{p^{2k+1}}.\label{rerek}
\end{eqnarray}
We note that $\mcal{T}_{\hbar}$ is to the order $\order{0}$. Now to complete our proof, we write
\begin{equation}\label{quantumme}
\mcal{T}_{\hbar}(q,p)=\sum_{k=0}^{\infty} (-1)^k \mcal{T}_k,
\end{equation}
where
\begin{equation}\label{45tr}
\mcal{T}_k=- \frac{(2k)!}{k!}\mu^{k+1}\sum_{j=0}^{k}\frac{\sigma_{k,j}}{2^j} b^{k-j} a^j \frac{q^{2k+1-j}}{p^{2k+1}}.
\end{equation}

To prove our assertion that $t_0(q,p)=\mcal{T}_{\hbar}(q,p)$, we need only to show that $\mcal{T}_k$ satisfies the initial value condition $t_0(q,p)$ and the recurrence relation is satisfied by the iterates of the local time of arrival at the origin. It is straightforward to show that $\mcal{T}_0(q,p)=t_0(q,p)$ by setting $k=0$ in equation (\ref{45tr}). It only remains to show that the $\mcal{T}_k$'s satisfy the recurrence relation (\ref{recur4xv}). Shifting index $k\rightarrow k-1$ in (\ref{45tr}) and substituting it back, together with the potential, in right hand side of equation (\ref{recur4xv}) yields,
\begin{eqnarray}
-\frac{\mu}{p}\int_0^q \frac{\partial V}{\partial q} \frac{\partial \mcal{T}_{k-1}}{\partial p} \, dq&=&-\frac{(2(k-1))!(2k-1)}{(k-1)!}\mu^{k+1} \nonumber\\
& & \times \sum_{j=0}^{k} \frac{1}{2^j} \frac{\left(2\sigma_{k-1,j-1}+\sigma_{k-1,j}\right)}{(2k+1-j)}\, b^{k-j}a^j \frac{q^{2k+1-j}}{p^{2k+1}}\label{45xyz},
\end{eqnarray}
where $\sigma_{k,j}=0$ for $0>j>k$. Equation (\ref{rerek}) converges to the local time of arrival at the origin if and only if equation (\ref{45xyz}) is equal to (\ref{45tr}) for all $k\geq 1$. Equating $\mcal{T}_k$ with equation (\ref{45xyz}), we find that $\sigma_{k,j}$ must the satisfy the following recurrence relation if strict equality is required,
\begin{equation}\label{lakoy}
\sigma_{k,j}=\frac{1}{(2k+1-j)}\left(\sigma_{k-1,j-1}+\frac{1}{2}\sigma_{k-1,j}\right).
\end{equation}
But this recurrence relation is already satisfied by the $\sigma_{k,j}$'s, as shown by equation (\ref{lakay}). Thus equation (\ref{rerek}) converges to the local time of arrival at the origin for the potential given.

We have thus proved what we have sought to prove that for linear systems, $\mcal{T}_{\hbar}(q,p)=t_0(q,p)$.

\subsection{Non-linear Systems}
In this section we show that for entire analytic potentials of the form
\begin{equation}\label{pote}
V(q)=\sum_{s=1}^{\infty}a_s\,q^s,
\end{equation}
with at least $a_3$ is non-vanishing, the proposed supraquantization of the time of arrival at the origin reduces only to the classical local time of arrival in the limit of vanishing or infinitesimal $\hbar$. Substituting the potential back into the time kernel equation and after some simplification, we arrive at the partial differential equation to solve
\begin{equation}\label{5potentialdiff}
	-2\frac{\hbar^2}{\mu}\frac{\partial^2 T}{\partial u \partial v} + \sum_{s=1}^{\infty}\frac{a_s}{2^{s-1}}\sum_{k=0}^{[s]}
			\binomial{2}{2k+1}\,u^{s-2k-1}\,v^{2k+1}\, T(u,v)=0
\end{equation}
where $[s]=\frac{s-1}{2}$ for $s$=odd and $[s]=\frac{s}{2}-1$ for $s$=even, subject to the same boundary conditions. 

Now let us assume an analytic solution of the form
\begin{equation}\label{etoraw}
T(u,v)=\sum_{m,n}\alpha_{m,n}\,u^m\, v^n,
\end{equation}
subject to the boundary conditions $\alpha_{m,n}=0$ for $m,n<0$, $\alpha_{m,0}=\frac{1}{4} \delta_{m,1}$, and $\alpha_{0,n}=0$ for all $n$. Substituting the assumed solution back into (\ref{5potentialdiff}) and collecting terms of equal powers of $u$ and $v$ yield the following recurrence relation among the coefficients,
\begin{equation}\label{nonlrecur}
\alpha_{m,n}=\frac{\mu}{2\hbar^2}\frac{1}{m\,n}\sum_{s=1}^{\infty}\frac{a_s}{2^{s-1}}\sum_{k=0}^{[s]}{s \choose 2k+1} \alpha_{m-s+2k,n-2k-2}
\end{equation}
Imposing the boundary condition, we have the following conditions imposed upon the coefficients: $\alpha_{1,0}=\frac{1}{4}$, $\alpha_{m,n}=0$ for all $n<0$ and $m\leq 1$, $\alpha_{m,0}=0$ for all $m\geq 2$, and $\alpha_{m,1}=0$ for all $m$. 

\subsubsection{Odd Powers of $v$}
The boundary conditions impose that the $\alpha_{m,n}$'s vanish for odd $n$ for all $m$. The coefficients already vanish for negative $n$, so we start with $n=1$. For $n=1$ equation  (\ref{nonlrecur}) gives
\begin{equation*}
\alpha_{m,1}=\frac{\mu}{2\hbar^2}\frac{1}{m\,\cdot 1}\sum_{s=1}^{\infty}\frac{a_s}{2^{s-1}}\sum_{k=0}^{[s]}{s \choose 2k+1} \alpha_{m-s+2k,-2k-1}.
\end{equation*}
Since $\alpha_{m,n}$ vanish for all negative $n$, $\alpha_{m,1}=0$ for all $m$. For $n=3$ we have
\begin{equation*}
\alpha_{m,3}=\frac{\mu}{2\hbar^2}\frac{1}{m\,\cdot 3}\sum_{s=1}^{\infty}\frac{a_s}{2^{s-1}}\sum_{k=0}^{[s]}{s \choose 2k+1} \alpha_{m-s+2k,1-2k}
\end{equation*}
Since $\alpha_{m,1}=0$ for all $m$ and $\alpha_{m,n}=0$ for all negative $n$, it follows that $\alpha_{m,3}=0$ for all $m$ as well. Now let $n=2j+1$, for $j=0,1,2,\dots$, and let $\alpha_{m,2j+1}=0$ for all $m$ for all $j\leq J$. Then for $n=2(J+1)+1$
\begin{equation*}
\alpha_{m,2(J+1)+1}=\frac{\mu}{2\hbar^2}\frac{1}{m\,(2(J+1)+1)}\sum_{s=1}^{\infty}\frac{a_s}{2^{s-1}}\sum_{k=0}^{[s]}{s \choose 2k+1} \alpha_{m-s+2k,2J+1-2k}.
\end{equation*} 
The $k=0$ term in the inner sum contains the factors $\alpha_{m-s+2k,2J+1}$, which are all vanishing because $\alpha_{m,2J+1}=0$ for all $m$; the $k=0$ term then does not contribute. The $k=1$ term contains the factors $\alpha_{m-s+2k,2(J-1)+1}$, which are all also vanishing because $\alpha_{m,2j+1}=0$ for all $m$ for all $j\leq J$; the $k=1$ term then does not contribute. 

Now for all $k\leq J$, the coefficients $\alpha_{m-s+2k,2J+1-2k}=\alpha_{m-s+2k,2(J-k)+1}$ vanish, again, because $\alpha_{m,2j+1}=0$ for all $m$ for all $j\leq J$; and no contribution comes from them. On the other hand, for all $k>J$, the coefficients $\alpha_{m-s+2k,2(J-k)+1}$ must vanish because of the condition that $\alpha_{m,n}=0$ for all negative $n$. Thus $\alpha_{m,2(J+1)+1}=0$ for all $m$ as well. We have already shown that $\alpha_{m,1}=0$ and $\alpha_{m,3}=0$ for all $m$, and it follows that $\alpha_{m,5}=0$ for all $m$ from what we have already shown, and so on. Thus it must be that $\alpha_{m,n}=0$ for all $m$ for every odd $n$. Odd powers of $v$ then do not contribute in the solution to the time kernel equation.

\subsubsection{Even Powers of $v$}
Now we proceed in determining the non-vanishing coefficients corresponding to even powers of $v$ or to even $n$. First, for $n=2$, the recurrence relation (\ref{nonlrecur}) reduces to
\begin{equation}\label{nln2}
	\alpha_{m,2}=\left(\frac{\mu}{2\hbar^2}\right) \frac{1}{m\cdot 2}
			\sum_{s=1}^{\infty} \frac{a_s}{2^{2-1}}
			\sum_{k=0}^{[s]} {s \choose 2k+1}\, \alpha_{m-2+2k,-2k}.
\end{equation}
Since the $\alpha_{m,n}$'s vanish for all negative $n$, only the $k=0$ term contributes in (\ref{nln2}). Thus
\begin{equation*}
	\alpha_{m,2}=\left(\frac{\mu}{2\hbar^2}\right) \frac{1}{m\cdot 2}
			\sum_{s=1}^{\infty} \frac{a_s}{2^{s-1}}\, {s \choose 1} \alpha_{m-s,0}.
\end{equation*}
But $\alpha_{m',0}=\frac{1}{4}\delta_{m',1}$, so that only the $s=m-1$ term contributes in the preceding relation. Since the power of $u$ is at least to the first order and $s\geq 1$, only those coefficients with $m\geq 2$ contribute above. Thus
\begin{equation}\label{memi}
	\alpha_{m,2}=\left(\frac{\mu}{2\hbar^2}\right) \frac{1}{m\cdot 2}
			\frac{a_{m-1}}{2^{m-2}} {m-1 \choose 1} \frac{1}{4}.
\end{equation}
The non-vanishing contributions from those with $m\geq 2$.

For $n=4$ the recurrence relation reduces to
\begin{equation}\label{nln4}
	\alpha_{m,4}=\left(\frac{\mu}{2\hbar^2}\right) \frac{1}{m\cdot 4}
			\sum_{s=1}^{\infty} \frac{a_s}{2^{s-1}}\sum_{k=0}^{[s]} \binomial{2}{2k+1}
			\, \alpha_{m-s+2k,2-2k}.
\end{equation}
For all $s$  only the $k=0,\, 1$ contribute to $\alpha_{m,4}$. All of the $s\geq 1$ terms contribute to the $k=0$ term. However, only those for $s\geq 3$ contribute to the $k=1$ term. Thus
\begin{equation*}
	\alpha_{m,4}=\left(\frac{\mu}{2\hbar^2}\right) \frac{1}{m\cdot 4}
				\sum_{s=1}^{\infty} \frac{a_s}{2^{s-1}} \binomial{s}{1} \alpha_{m-s,2}
				+ \left(\frac{\mu}{2\hbar^2}\right) \frac{1}{m\cdot 4}
				\sum_{s=3}^{\infty} \frac{a_s}{2^{s-1}} \binomial{s}{3} \alpha_{m-s+2,0}
\end{equation*}		
Since $\alpha_{m',2}$ is non-vanishing only for $m'\geq 2$, it has to be that $(m-s)\geq 2$ in the first term; thus only those $1\leq s\leq (m-2)$ contribute in the sum. Since $s\geq 1$ only those with $m\geq 3$ contribute. On the other hand, only $s=m+1$ contributes in the second term. Thus, upon substituting $\alpha_{m',2}$ and $\alpha_{m',0}$,
\begin{eqnarray*}
	\alpha_{m,4}&=& \left(\frac{\mu}{2\hbar^2}\right)^2 \frac{1}{m\cdot 4}
				\sum_{s=1}^{m-2} \frac{a_s\, a_{m-s-1}}{2^{m-2}} 				   \binomial{s}{1}\binomial{m-s-1}{1}\frac{1}{4} \nonumber \\
				& &+ \left(\frac{\mu}{2\hbar^2}\right) \frac{1}{m\cdot 4}
					\frac{a_{m+1}}{2^{m}} \binomial{m+1}{3} \frac{1}{4},
\end{eqnarray*}
with the first term having contribution only for $m\geq 3$ and the second term for all $m\geq 2$. 

We would like now to generalize our results for arbitrary $j$. The explicit forms of $\alpha_{m,0}$, $\alpha_{m,2}$ and $\alpha_{m,4}$ suggest that, for some fixed $j$, we have
\begin{equation}\label{bokbok}
\alpha_{m,2j}=\sum_{s=0}^{j-1} \alpha_{m,j}^{(s)} \left(\frac{\mu}{2\hbar^2}\right)^{j-s},
\end{equation}
for some constants $\alpha_{m,j}^{(0)}$ independent of $\hbar$ and $\mu$. We prove (\ref{bokbok}) by induction and consequently determine the recurrence relation satisfied by these constants that determines them uniquely.  Now for $n=2j$ for some $j\geq 1$, the recurrence relation (\ref{nonlrecur}) can be written in the form
\begin{equation}\label{nonlrecurmod}
\alpha_{m,2j}=\frac{\mu}{2\hbar^2}\frac{1}{m\cdot 2j}\sum_{s=1}^{\infty}\frac{a_s}{2^{s-1}}\sum_{k=0}^{j-1}{s \choose 2k+1} \alpha_{m-s+2k,2(j-k-1)}.
\end{equation}
We have replaced $[s]$ with $j-1$ in the inner summation limit because whatever extra terms are introduced they are taken care of by the binomial factor. The order of summation can be reordered to yield
\begin{eqnarray}\label{nonlmod}
\alpha_{m,2j}&=&\frac{\mu}{2\hbar^2}\frac{1}{m\cdot 2j}\sum_{k=0}^{j-1}\sum_{s=2k+1}^{\infty}\frac{a_s}{2^{s-1}} {s \choose 2k+1} \alpha_{m-s+2k,2(j-k-1)}\nonumber\\
&=&\frac{\mu}{2\hbar^2}\frac{1}{m\cdot 2j}\sum_{k=0}^{j-1}\sum_{s=2k+1}^{m+2k-1}\frac{a_s}{2^{s-1}} {s \choose 2k+1} \alpha_{m-s+2k,2(j-k-1)}
\end{eqnarray}
The second line follows from the fact that $\alpha_{m-s+2k,2(j-k-1)}$ is non-vanishing only when $m-s+2k\geq 1$ or $m+2k-1\geq s$. 

Now we substitute equation (\ref{bokbok}) back into the right hand side of equation (\ref{nonlmod}). This yields
\begin{equation}
\alpha_{m,2j}=\frac{1}{m\cdot 2j} \sum_{k=0}^{j-1}\sum_{r=0}^{j-k-1} \left(\sum_{s=2k+1}^{m+2k-1}\frac{a_s}{2^{s-1}}{s \choose 2k+1} \alpha_{m-s+2k,j-k-1}^{(r)}\right)\, \left(\frac{\mu}{2\hbar^2}\right)^{j-k-r}\label{oko}
\end{equation}
We can rearrange equation (\ref{oko}) to obtain the following simplification,
\begin{equation}
\alpha_{m,2j}=\frac{1}{m\cdot 2j} \sum_{s=0}^{j-1}\left(\sum_{r=0}^{s} \sum_{l=2r+1}^{m+2r-1}\frac{a_l}{2^{l-1}}{l \choose 2r+1} \alpha_{m-l+2r,j-r-1}^{(s-r)}\right) \left(\frac{\mu}{2\hbar^2}\right)^{j-s}.\label{gog1}
\end{equation}
Expression (\ref{bokbok}) holds if and only if it equals the right hand side of (\ref{gog1}) for all $j$. Strict equality holds if and only if the constants $\alpha_{m,j}^{(s)}$ satisfy the recurrence relation
\begin{equation}
\alpha_{m,j}^{(s)}=\frac{1}{m\cdot 2j} \sum_{r=0}^{s} \sum_{l=2r+1}^{m+2r-1}\frac{a_l}{2^{l-1}}{l \choose 2r+1} \alpha_{m-l+2r,j-r-1}^{(s-r)},\label{kikokiko}
\end{equation}
for all $0\leq s \leq (j-1)$, subject to the initial value $\alpha_{m,0}^{(0)}=\frac{1}{4}\delta_{m,1}$. Equations (\ref{oko}) and (\ref{kikokiko}) now define the non-vanishing coefficients for even powers of $v$. They can be solved explicitly. In the following we are only interested in the classical limit.

\subsection{The Classical Coefficients}
We now identify the contributing coefficients and determine the leading order of $\hbar$ correction in the classical limit. For a fixed even $n=2j$, $j=0, 1, 2, \dots$, we have seen above that the contributing coefficients $\alpha_{m,2j}$'s are of the form
\begin{equation*}
\alpha_{m,2j}=\sum_{s=0}^{j-1}\left(\frac{\mu}{2}\right)^{j-s} \frac{\alpha_{m,j}^{(s)}}{\hbar^{2(j-s)}}.
\end{equation*}
The contribution of each term is $\alpha_{m,2j} v^{2j}$ (the $u^m$ factor is left out because it is not relevant in determining the $\hbar$-order of contribution in the classical limit). The classical contribution of this term is proportional to
\begin{eqnarray}
\frac{1}{\hbar}\alpha_{m,n}\int_{-\infty}^{\infty} \mbox{sgn}(v)\,v^{2j}\, \exp\left(-i\frac{vp}{\hbar}\right)\, dv 
&\propto& \sum_{s=0}^{j-1}\left(\frac{\mu}{2}\right)^{j-s} {\alpha_{m,j}^{(s)}}\hbar^{2s},\label{u56}
\end{eqnarray}
where we have arrived at the left hand side of the first line by using the prescribed classical transition and with the right hand using identity (\ref{identity}). We see immediately that the only contributing term in the classical limit corresponds to $s=0$. Moreover, we can already see that the leading $\hbar$ correction in the classical limit is $\order{2}$; this corresponds to $s=1$ in equation (\ref{u56}).

Thus the coefficients contributing only in the classical limit corresponds to those  for $s=0$ for a given $j$. 
And these coefficients satisfy the recurrence relation,
\begin{equation}\label{recurCC}
\lterm{m}{2j}=\frac{1}{m\cdot 2j}\sum_{s=1}^{\infty}\frac{s\,a_s}{2^{s-1}}\lterm{m-s}{j-1}.
\end{equation}
With $\alpha_{m,0}=\frac{1}{4}\delta_{m,1}$, we can generate the following first few coefficients
\begin{eqnarray*}
\alpha_{m,0}^{(0)}&=& \frac{1}{4}\delta_{m,1}\\
\\
\lterm{m}{1}&=&\frac{1}{2\,m}\sum_{s=1}^{\infty}\frac{s\,a_s}{2^{s-1}}\lterm{m-s}{0}\\
&=&\frac{1}{1\cdot\,m}\frac{(m-1)\,a_{m-1}}{2^{m+1}}\;\;\; \mbox{for all}\;\;\; m\geq 2 \\
\\
\alpha_{m,2}^{(0)}&=&\frac{1}{2\,m}\sum_{s=1}^{\infty}\frac{s\,a_s}{2^{s-1}}\alpha_{m-s,1}\\
&=&\frac{1}{1\cdot2\,m\,2^{m+1}}\sum_{s=1}^{m-2}\frac{s\,a_s}{(m-s)}\left(m-s-1\right)\,a_{m-s-1},\;\;\;\mbox{for all}\;\;\; m\geq 3 \\
\end{eqnarray*}
From these few iterations, we infer that the coefficients are given by
\begin{equation}\label{cococo}
\lterm{m}{j}= \frac{C_{m,j}}{j!\cdot2^{m+1}\,m}.
\end{equation}
where the $C_{m,j}$'s are constants, for all $m\geq(j+1)$. Substituting this expression back into the recurrence relation (\ref{recurCC}), yields the recurrence relation satisfied by the $C_{m,j}$'s,
\begin{equation}
C_{m,j}=\sum_{s=1}^{m-j} \frac{s\, a_s}{(m-s)}C_{m-s,j-1}.
\end{equation}
This is uniquely solved by specifying the initial value. Setting $j=0$ in equation (\ref{cococo}) and comparing it with the known value of $\alpha_{m,0}^{(0)}$ yields the initial value $C_{m,0}=\delta_{m,1}$. The recurrence relation can be solved explicitly, but we do not need to write it down.

\subsection{The Solution}
The coefficients $\alpha_{m,j}^{(0)}$ give the group of contributions with the order $\order{0}$ in the classical limit. For every $j\geq 0$, there is a contribution $T_j (u,v)$'s in the solution $T(u,v)$, which is given by
\begin{eqnarray*}
T_j (u,v)= \left(\frac{\mu}{2\hbar^2}\right)^j\sum_{m=j+1}^{\infty} \alpha_{m,2j}^{(0)} u^m\, v^{2j}=\left(\frac{\mu}{2\hbar^2}\right)^j \frac{1}{j!}\sum_{m=j+1}^{\infty} \frac{C_{[m,j]}}{m\cdot 2^{m+1}}u^m\,v^{2j}.
\end{eqnarray*}
The solution to (\ref{newdequation}) can then be written in the form
\begin{equation}
T(u,v)=\sum_{j=0}^{\infty}T_j (u,v) + S(u,v),
\end{equation}
where  the second term is responsible for order $\order{2}$ in the classical limit. The solution to the time kernel equation in the $(q,q')$ coordinate then assumes the form
\begin{equation}
T(q,q')=\sum_{j=0}^{\infty}T_j (q,q') + S(q,q'),
\end{equation}
in which $T_j (q,q')$ derives from $T_j (u,v)$ with the substitutions $u=(q+q')$ and $v=(q-q')$.

The functional kernel of the generalized time of arrival operator then splits in two parts
\begin{eqnarray*}
\kernel{\mcal{T}}=\sum_{j=0}^{\infty}\kernel{\opr{T}_j}\,+\, \kernel{\Delta\mcal{T}} \label{kernelsplit}
\end{eqnarray*}
where $\kernel{\opr{T}_j}=\frac{\mu}{i \hbar}\, T_j (q,q')\, \mbox{sgn}(q-q')$. Each of this $\kernel{\opr{T}_j}$ contributes in the classical limit,
\begin{eqnarray*}
T_j(q,p)&=&  \int_{-\infty}^{\infty}\transker[T_j]\, \expo{\left(-\frac{i\,p\, v}{\hbar}\right)}\, dv\nonumber \\
&=&-\, (2j-1)!!(-1)^j \;\frac{\mu^{j+1}}{p^{2j+1}}\sum_{m=j+1}^{\infty}\frac{C_{m,j}}{m} q^{m},
\end{eqnarray*}
where a simplification has been made in the second line.

To prove that $\sum_{j=0}^{\infty} T_j(q,p)$ converges to the local time of arrival, we write the term with leading order $\order{0}$ in the form
\begin{equation}
\mcal{T}_{\hbar}^0 (q,p)=\sum_{k=0}^{\infty}(-1)^k \mcal{T}_k (q,p).
\end{equation}
\begin{equation}\label{oo0}
\mcal{T}_k=-(2k-1)!! \;\frac{\mu^{k+1}}{p^{2k+1}}\sum_{m=k+1}^{\infty}\frac{C_{m,k}}{m} q^{m}.
\end{equation}
$\mcal{T}_{\hbar}^0$ converges to the local time of arrival in the origin if the $\mcal{T}_0$ reproduces the initial value and the the remaining terms satisfy the recurrence relation for the local time of arrival. Since $C_{m,0}=\delta_{m,1}$, for $k=0$, we have
\begin{equation}
\mcal{T}_0 (q,p)=-\mu \frac{q}{p},
\end{equation}
as required. 

It remains to show that the rest of the terms satisfy the recurrence relation (\ref{recur4xv}). Shifting index $k\rightarrow (k-1)$ in $\mcal{T}_{k}$, we have
\begin{eqnarray}
\frac{\partial V}{\partial q}\frac{\partial \mcal{T}_{k-1}}{\partial p}&=&(2k-1)!! \frac{\mu^k}{p^{2k}}\sum_{s=1}^{\infty} s\, a_s q^{s-1}\cdot \sum_{m=k}^{\infty} \frac{C_{m,k-1}}{m}q^m\nonumber\\
&=& (2k-1)!! \sum_{k+1}^{\infty} \sum_{r=1}^{s-r} \frac{r\, a_r C_{s-r,k-1}}{(s-r)} q^{s-1},
\end{eqnarray}
where we have used the identity
\begin{equation*}
\sum_{s=1}^{\infty} a_s x^s \cdot \sum_{k=m}^{\infty} b_k x^k = \sum_{l=m+1}^{\infty} \sum_{n=1}^{l-m} a_n \, b_{l-n} x^l
\end{equation*}
to arrive at the second line. Now we have
\begin{equation}\label{openg}
-\frac{\mu}{p}\int_0^q dq\, \frac{\partial V}{\partial q}\frac{\partial \mcal{T}_{k-1}}{\partial p}
=(2k-1)!! \sum_{k+1}^{\infty} \sum_{r=1}^{s-r} \frac{r\, a_r C_{s-r,k-1}}{(s-r)} \frac{q^s}{s}.
\end{equation}
$\mcal{T}_{\hbar}$ converges to the local time of arrival at the origin if and only if equations (\ref{oo0}) and (\ref{openg}) are strictly equal for all $k$. Equating them, we find that strict equality for all $k$ holds if and only if the $C_{m,k}$'s satisfy the recurrence relation
\begin{equation}
C_{m,k}=\sum_{r=1}^{m-k} \frac{r\,a_r}{(m-r)}C_{m-r,k-1}.
\end{equation}
But this is just the recurrence relation we have arrived at above. Thus the $\mcal{T}_k$'s satisfy the initial value condition and the required recurrence relation. The leading term $\mcal{T}_{\hbar}^0$ then converges to the local time of arrival at the origin, as what we have sought to prove.

\subsection{Integral Form of the Classical Term}
In the above discussion, we did not bother to consider the convergence of the group of terms contributing in the classical limit, the terms with $\order{0}$ when Weyl-Wigner transformed; we denote this group of terms by $\kernel{\mcal{T}_0}$, and call it the classical term. Here we show that $\kernel{\mcal{T}_0}$ converges everywhere in the $qq'$-plane. We do this by showing that it has an integral representation which is defined everywhere. 

It will be sufficient for us to derive the integral form of the time kernel for the linear case, because the non-linear case can be derived similarly. Our goal is to rewrite equation (\ref{dung}) such that it is explicitly everywhere convergent in the $qq'$-plane. We do this as follows. In equation (\ref{45tr}), we have the following expression for $\mcal{T}_k$,
\begin{equation}\label{re45tr}
\mcal{T}_k=- \frac{(2k)!}{k!}\frac{\mu^{k+1}}{p^{2k+1}}\sum_{j=0}^{k}\frac{\sigma_{k,j}}{2^j} b^{k-j} a^j {q^{2k+1-j}}.
\end{equation}
We compare this with equation (\ref{okay}) for $x=0$,
\begin{equation}\label{okayrepeat}
	T_k =-\frac{(2k-1)!!}{k!} \frac{\mu^{k+1}}{p^{2k+1}}\int_0^q \left(V(q)-V(q')\right)^k\, dq'.
\end{equation}

Since we already know that equation (\ref{quantumme}) converges absolutely to the local time of arrival at the origin, it must be that equations (\ref{re45tr}) and (\ref{okayrepeat}) are equal for all $k$. Equating them and changing variables $q\rightarrow \frac{1}{2}(q+q')$ in the resulting equality gives us the following identity
\begin{equation}\label{adinggg}
\sum_{j=0}^{k} \sigma_{k,j} b^{k-j} a^j (q+q')^{2k+1-j}=2^{2k+1}\frac{(2k-1)!}{(2k)!}\left.\int_0^s\left(V(s)-V(q'')\right)^k\, dq''\right|_{s=\frac{1}{2}(q+q')},
\end{equation}
which is the simplification we need in equation (\ref{dung}). Substituting equation (\ref{adinggg}) into equation (\ref{dung}) yields
\begin{eqnarray}
T(q,q')&=&\frac{1}{2}\sum_{k=0}^{\infty} \left(\frac{\mu}{\hbar^2}\right)^k\frac{(2k-1)!!}{(2k)!k!} (q-q')^{2k} \left.\int_0^s\left(V(s)-V(q'')\right)^k\, dq''\right|_{s=\frac{1}{2}(q+q')}\nonumber\\
&=&\frac{1}{2}\left.\int_0^s dq''\, _{0}F_1\left(1;\left(\frac{\mu}{2\hbar^2}\right)(q-q')^2\left(V(s)-V(q'')\right)\right)\right|_{s=\frac{1}{2}(q+q')},
\end{eqnarray}
where $_{p}F_{q}$ is the generalized hypergeometric function. The integration can be pulled out of the summation because of the continuity of the potential and the absolute everywhere convergence of the hypergeometric function for $p<q$.  $T(q,q')$ is consequently defined everywhere. Finally the time kernel is explicitly given by
\begin{equation}\label{classic}
\kernel{\mcal{T}_0}=\frac{\mbox{sgn}(q-q')}{2i\hbar} \int_0^{\frac{1}{2}(a+q')}\!\!dq''\, _{0}F_{1}\! \left(1; \frac{\mu}{2\hbar^2}(q-q')^2\left\{V\left(\frac{1}{2}(q+q')\right)-V\left(q''\right)\right\}\right),
\end{equation}
We have arrived at equation (\ref{classic}) for linear systems, but similar working on the classical term of the solution for the non-linear case yields the same expression (\ref{classic}), in which $V(q)$ is now the appropriate potential for non-linear systems.

Since the time kernel and the classical term coincide for linear systems, the time kernel is defined everywhere, and thus a functional in $\Phi^{\times}$, and it defines a generalized observable relative to the rigging provided by $\Phi$ (see Appendix). For non-linear systems, the leading term is likewise defined everywhere, and it defines a generalized observable relative to $\Phi$. We have not been able to investigate the functional structure of the remaining terms for non-linear systems. Generally the time kernels for entire analytic potentials can then be written in the form
\begin{eqnarray}
\mbox{Linear Systems:}& &\kernel{\mcal{T}}=\kernel{\mcal{T}_0}\label{566y}\\
\mbox{Non-Linear Systems:}& &\kernel{\mcal{T}}=\kernel{\mcal{T}_0}+\kernel{\Delta\mcal{T}}\label{566x}.
\end{eqnarray}

Comparison of $\kernel{\mcal{T}_0}$ with the Weyl quantization of the local time of arrival in the origin shows that they are equal. One can check this for himself by applying Weyl's quantization prescription (\ref{kerker}) to the local time of arrival. Weyl quantization then agrees only with the result of supraquantization for linear systems and it fails to satisfy the required commutator value for non-linear systems. By our results for the non-linear system the second term in equation (\ref{566x}) is to the $\order{2}$ in the classical limit.

\section{Supraquantization for Arbitrary Points of\\Arrival $x$}\label{arbitrary}
Having solved the time of arrival supraquantization problem at the origin, now we show that our results above can be imported to solve the supraquantization at an arbitrary point $x$. Generally the classical time of arrival at a point $x$ is given by equation (\ref{classpass}).
Changing variables in equation (\ref{classpass}) to $\left(\tilde{q}=q-x,\tilde{p}=p\right)$, the expression for the time of arrival reduces to
\begin{equation}\label{classpass12}
		T_x(\tilde{q},\tilde{p})=-\sign{\tilde{p}} \sqrt{\frac{\mu}{2}}\int_0^{\tilde{q}} \frac{d\tilde{q}'}{\sqrt{H(\tilde{q}+x,\tilde{p})-V(\tilde{q}'+x)}}
\end{equation}
Comparing equation (\ref{classpass12}) with the classical time of arrival at the origin, we find that the expression is equivalent to the time of arrival at the origin under the potential $\tilde{V}(\tilde{q})=V(\tilde{q}+x)$.

The surpraquantization for arbitrary arrival points $x$ then can be solved by solving the time kernel equation at the origin subject to the potential $\acute{V}(\acute{q})=V(\acute{q}+x)$. For this case the time kernel equation assumes the form
\begin{equation*} 
 -\frac{\hbar^2}{2\mu}\frac{\partial^2 T_x(\tilde{q},{\tilde{q}}')}{\partial q^2}+\frac{\hbar^2}{2\mu} \frac{\partial^2 T_x(\tilde{q},{\tilde{q}}')}{\partial {{{\tilde{q'}}}}^2}+ \left(V(\tilde{q}+x)-V(\tilde{q}'+x)\right)T_x(\tilde{q},{\tilde{q}}') =0
\end{equation*}
and the solution is still subject to the same boundary conditions
\begin{equation}
		T_x(\tilde{q},\tilde{q})=\frac{\tilde{q}}{2},\;\;\;
		T_x(\tilde{q},-\tilde{q})=0.
\end{equation}
After solving for $T_x(\tilde{q},\tilde{q}')$, we can transform back to the original coordinate to get the kernel for the original problem. And that completes the supraquantization of the classical time of arrival for arbitrary $x$. Note that our earlier result in the neighborhood of the origin is subsumed in the above solution by simply setting $x=0$.

\section{Discussion and Conclusion}\label{conclusion}

In this paper we have demonstrated that the classical time of arrival can be derived  quantum mechanically without solving for and inverting the classical equations of motion. Our results, albeit still needing more clarifications (especially in the non-linear case), undoubtedly forces us to reconsider our ideas on quantization, and consider supraquantization in places where quantization fails.  Generally it is known that obstruction to quantization exists, so that no quantization is possible to consistently satisfy the required commutation relations. For example Weyl quantization cannot consistently quantize all classical observables as we have demonstrated for the class of time of arrival observables. What is generally done is to choose an elite class of the classical observable that can be consistently quantized and derive the rest of the quantum observables by expressing them in terms of this elite class. In Euclidean space, the choice is usually the Heisenberg class, the position and momentum operators, together with the identity operator. The rest of the quantum observables are then derived by expressing them in terms of this class of operators. This, however, is not wholly satisfactory because the resulting operators do not necessarily satisfy the required algebra.

Now if we strongly require consistency with the required algebra of observables in spaces where obstruction to quantization exists, then we must leave quantization and find an alternative platform. It is here that the idea of supraquantization may come in. However, its implementation may not be straightforward. As what we have discussed earlier, supraquantization may necessarily require some classification of observables, as opposed to quantization which does not classify observables. The classification is necessary, at least for the class of time of arrival observables, in identifying the characteristic properties of the class that can be used in implementing the transfer principle. The natural questions are {\it How do we get the appropriate classification and how do we identify the characteristic properties of the class}? These may not be easily answered, but they will eventually require us to go back to the basic definition of the elements of the class and the appropriate axioms of quantum mechanics to impose on them. 

Assuming that we have settled the first question, we may use quantization itself as a tool in addressing the second question. What we can do is the following: Given a class $\mcal{C}$ of classical observables, divide $\mcal{C}$ in two parts $\mcal{C}_N$ and $\mcal{C}_O$. The subclass $\mcal{C}_N$, which we may call the non-obstructed class, consists of those observables that can be consistently quantized; and the subclass $\mcal{C}_O$, which we may call the obstructed class, consists of those observables that can not be consistently quantized. We can work on the $\mcal{C}_N$ using quantization and determine the properties that can be extended to the rest of the class. Once the common property of all those in $\mcal{C}_N$ has been determined, one can use the transfer principle in treating the obstructed class $\mcal{C}_O$. For the class of (classical) time of arrival observables, we find that the non-obstructed class with respect to Weyl quantization consists of all linear systems, while the obstructed class consists of all non-linear systems. Following the above suggestion, we could have arrived at the same solution by working directly with the linear system and extending the result to nonlinear systems via the appropriate transfer principle. The example of the classical time of arrival demonstrates how obstruction to quantization can be  formally circumvented with the idea of supraquantization. 

\renewcommand{\theequation}{A-\arabic{equation}}
\setcounter{equation}{0}  
\setcounter{section}{0}
\section{Appendix}
To establish and to avoid possible confusion with our notation for the RHS-extensions and RHS-reductions, in particular the use of the notation $\opr{A}_{\times}\varphi=\dirac{F_{\opr{A}}}{\varphi}=\opr{A}\varphi$, we give an example.

 Consider the momentum operator $\opr{P}$ in the Hilbert space $\mathcal{H}=L^2(\Re,dq)$. The domain of $\opr{P}$, $\mathcal{D}(\opr{P})$, consists of all vectors in $\mcal{H}$ that are almost differentiable everywhere in the real line, and whose first derivatives are Lesbegue square integrable. For every vector $\varphi(q)$ in $\mathcal{D}(\opr{P})$, the momentom operator acts as $\left(\opr{P}\varphi\right)\!(q)=-i\hbar\varphi'(q)$. By definition $\opr{P}$ is self-adjoint so that $\opr{P}=\opr{P}^{\dagger}$. Now we choose the rigging $\Phi^{\times}\supset \mcal{H} \supset \Phi$, where $\Phi$ is the space of infinitely differentiable complex valued functions with compact support in the real line, and $\Phi^{\times}$ its corresponding functional space. Since $\Phi$ is contained in $\mathcal{D}(\opr{P})$, we can define its rigged Hilbert space extension and reduction. For every vector $\varphi$ in $\Phi$ and functional $\phi$ in $\Phi^{\times}$, we can write $\dirac{\phi}{\varphi}=\int_{\Re}\phi^*(q)\varphi(q)\, dq$, where the integration is understood in the distributional sense when singular $\phi$ is involved, say, the Dirac delta. 

Now the RHS-extension of $\opr{P}$ is found as follows: For every $\varphi$ in $\Phi$ and $\phi$ in $\Phi^{\times}$, we have 
\begin{eqnarray}
\dirac{\phi}{\opr{P}^{\dagger}\varphi}&=&\int_{\Re}\phi^*(q)\left(-i\hbar\varphi'(q)\right)dq\nonumber \\ 
&=&\int_{\Re}i\hbar \phi'^*(q)\varphi(q)\, dq\nonumber\\
&=&\int_{\Re}\left(-i\hbar\phi'(q)\right)^*\varphi(q)\, dq\nonumber\\
&=&\int_{\Re}\left(\opr{P}^{\times}\phi\right)\!\!(q)\varphi(q)\, dq\nonumber,
\end{eqnarray}
where the second line follows from the definition of the derivatives of functionals. The RHS-extension of $\opr{P}$ is then given by the operator $\opr{P}^{\times}$ which acts everywhere in $\Phi^{\times}$ as $\opr{P}^{\times}\phi=-i\hbar\phi'$.

On the other hand the RHS-reduction of $\opr{P}$ is found as follows: First, we have to indicate the reduction of $\opr{P}$ in $\Phi$. Its reduction is simply the operator $\opr{P}_{\Phi}$, which acts only on vectors $\varphi$ in $\Phi$ according to $\left(\opr{P}_{\Phi}\varphi\right)\!\!(q)=-i\hbar\varphi'(q)$. Second, we have to find the functional $F_{\opr{P}}(q)$ in $\Phi^{\times}$ for every $q$ in the real line, such that $\dirac{F_{\opr{P}}(q)}{\varphi}=-i\hbar\varphi'(q)$, for all $\varphi$ in $\Phi$. By inspection, this functional is given by $F_{\opr{P}}(q)=-i\hbar\delta'(q-q')$. It is so because
\begin{eqnarray*}
\dirac{F_{\opr{P}}(q)}{\varphi}&=&\int_{\Re}F_{\opr{P}}(q)^*\varphi(q')\, dq'\\
&=&\int_{\Re}i\hbar \delta'(q-q')\varphi(q')\, dq'\\
&=&\int_{\Re}\delta(q-q')\left(-i\hbar\varphi'(q')\right)\,dq'\\
&=&-i\hbar\varphi'(q)\\
&=&\left(\opr{P}_{\Phi}\right)\!\!(q).
\end{eqnarray*}
Thus, by our definition, the uniquely associated functional to $\opr{P}$ is the functional $-i\hbar\delta'(q-q')$. The RHS-reduction of $\opr{P}$ is now symbolically given by
\begin{equation}
\opr{P}_{\times}=\dirac{F_{\opr{P}}}{\cdot}=\int_{\Re}dq'\,i\hbar\delta'(q-q')\nonumber,
\end{equation}
with $F_{\opr{P}}^*=i\hbar\delta(q-q')$ as the functional kernel of $\opr{P}_{\times}$.

Note that possible confusion may arise when the above notation is used, for example, in expressions like $\dirac{F(\varphi)}{\varphi}=\dirac{\dirac{F}{\varphi}}{\varphi}$, such as in the definition of generalized observables. The confusion may creep in when one interprets $\dirac{F}{\varphi}$ as a constant scalar number. While $\dirac{F}{\varphi}$ is indeed a scalar number, it may be understood to range in the complex plane, such as $\dirac{F_{\opr{P}}(q)}{\varphi}$ in the above example, so that $\dirac{F}{\varphi}$ can be understood as a vector in $\Phi$ or $\Phi^{\times}$, whichever the case maybe.

\section{Appendix}  
Let us consider the function 
\begin{equation}\label{apensum}
F(q,q')=\mbox{sgn}(q-q')\sum_{k=0}^{\infty}T_k(q,q')
\end{equation}
where the summation is everywhere absolutely convergent or entire analytic in the $qq'$-plane. Now for a fixed $q$, is $F(q,q')$ a functional belonging to $\Phi^{\times}$? 

First it has to be that for all $\varphi$ in $\Phi$, $\abs{\dirac{F}{\varphi}}<\infty$. Let us denote the sum in equation (\ref{apensum}) by $S(q,q')$. Then for all $\varphi$ in $\Phi$
\begin{equation}
\abs{\int_{\Sigma}F(q,q')\varphi(q')dq'}\leq \mbox{sup}_{\Sigma}\abs{S(q,q')}\abs{\int_{\Sigma}{\varphi(q)}dq},
\end{equation}
where $\Sigma$ is the support of $\varphi(q)$. The right hand side of the above inequality is finite because $S(q,q')$ is bounded in any finite region of the $qq'$-plane. Second it has to be that for every sequence $\varphi_n$ in $\Phi$ converging to zero in $\Phi$, $\dirac{F}{\varphi_n}$ converges to $0$. This follows immediately because $F(q,q')$ is locally integrable. Thus $\kernel{\mcal{T}}$ is a functional belonging to $\Phi^{\times}$ for a fixed $q$. 

Now for arbitrary $\varphi$ in $\Phi$, is $G(q)=\int F(q,q')\varphi(q')dq'$ a functional belonging to $\Phi^{\times}$? For all $\phi(q)\in \Phi$, 
\begin{equation}\label{apensumo}
\abs{\int_{\Sigma'}\int_{\Sigma}F(q,q')\phi(q)^*\varphi(q')dq'\,dq}\leq \mbox{sup}_{\Sigma\times\Sigma'}\abs{S(q,q')}\abs{\int_{\Sigma\times\Sigma'}{\phi^*(q) \varphi(q')}dq\,dq'}.
\end{equation}
The right hand side of the inequality is finite because $S(q,q')$ is bounded in every bounded region of the $qq'$-plane. Now it is sufficient to show that for every sequence $\phi_n$ converging to zero in $\Phi$, $\dirac{G}{\phi_n}\rightarrow 0$. This follows immediately by substituting $\phi_n$ in inequality (\ref{apensumo}) for $\phi$. 

Thus $F(q,q')$ is the functional kernel of an operator $\mcal{F}:\Phi\mapsto \Phi^{\times}$ and thus $\dirac{F}{\varphi}$ is itself a functional in $\Phi^{\times}$. 

\section*{Acknowledgement}
The author acknowledges the many fruitful discussions with H. Domingo which led to numerous refinements of the original manuscript, and acknowledges R. de la Madrid for his insightful remarks on rigged Hilbert space theory; and likewise thanks I. Tagaca for her careful reading and suggetions to improve the textual presentation of the entire manuscript.

\end{document}